\documentclass[aps,pra,twocolumn,export]{revtex4-1} 
\usepackage[colorlinks=true,citecolor=blue, urlcolor=blue, linkcolor=blue]{hyperref}

\usepackage[utf8]{inputenc}
\usepackage[german, english]{babel}
\usepackage{comment}
\usepackage{graphicx}
\usepackage{amssymb}
\usepackage{epstopdf}
\usepackage{graphicx} 
\usepackage{dcolumn}   % needed for some tables
\usepackage{ulem}
\usepackage{bm}        % for math
\usepackage{tikz}
\usepackage{amssymb}   % for math
\usepackage{amsmath}   % for math
\usepackage{amsthm}
\usepackage{dsfont}
\usepackage{mathtools}
\usepackage{mathrsfs}   
\usetikzlibrary{matrix,arrows.meta,decorations.markings,calc}
\usetikzlibrary{decorations.pathreplacing}
\usepackage{tcolorbox}
\usepackage{subfigure}
\usepackage{soul}
\usepackage{xspace}

\newcommand{\id}{\mathds{1}}
\newcommand{\tr}{\operatorname{Tr}}
\newcommand{\swap}{\scalebox{0.8}{\sf SWAP}\xspace}
\usepackage{cleveref}

\usetikzlibrary{3d}
\usetikzlibrary{decorations.pathreplacing,decorations.markings,calc}

\newcommand{\myinv}[1]{#1^{\scalebox{0.9}[1.0]{-}1}}

\usepackage{color}

\tikzset{
  every picture/.style = {
    baseline={([yshift=-.5ex]current bounding box.center)}, 
    scale=1.2,
    transform shape,
    font=\scriptsize
  }
}

\tikzset{
  3dpeps/.pic={
    \begin{scope}[canvas is zx plane at y=0]
      \draw (-0.5,0)--(0.5,0);
      \draw (0,-0.4)--(0,0.4);
      \filldraw (0,0) circle (0.07);
    \end{scope}
    \draw (0,0,0)--(0,0.2,0);
    
  }
}

\tikzset{
  3dpepsdown/.pic={
    \draw (0,0,0)--(0,-0.2,0);
    \begin{scope}[canvas is zx plane at y=0]
      \draw (-0.5,0)--(0.5,0);
      \draw (0,-0.4)--(0,0.4);
      \filldraw (0,0) circle (0.07);
    \end{scope} 
  }
}

\tikzset{
  3dpepsshort/.pic={
    \begin{scope}[canvas is zx plane at y=0]
      \draw (-0.5,0)--(0.5,0);
      \draw (0,-0.3)--(0,0.3);
      \filldraw (0,0) circle (0.07);
    \end{scope}
    \draw (0,0,0)--(0,0.2,0);
    
  }
}

\tikzset{
  3dpepsdownshort/.pic={
    \draw (0,0,0)--(0,-0.2,0);
    \begin{scope}[canvas is zx plane at y=0]
      \draw (-0.5,0)--(0.5,0);
      \draw (0,-0.3)--(0,0.3);
      \filldraw (0,0) circle (0.07);
    \end{scope} 
  }
}

 \tikzset{
  3dpepsp/.pic={
    \begin{scope}[canvas is zx plane at y=0]
      \draw (-0.3,0)--(0.3,0);
      \draw (0,-0.25)--(0,0.25);
      \filldraw[draw=black,fill=blue] (0,0) circle (0.07);
    \end{scope}
  }
}

 \tikzset{
  3dpepspb/.pic={
    \begin{scope}[canvas is zx plane at y=0]
      \draw (-0.3,0)--(0.3,0);
      \draw (0,-0.25)--(0,0.25);
      \filldraw[draw=black,fill=black] (0,0) circle (0.07);
    \end{scope}
  }
}

\tikzset{
  3dpepsdownpb/.pic={
    \begin{scope}[canvas is zx plane at y=0]
      \draw (-0.3,0)--(0.3,0);
      \draw (0,-0.25)--(0,0.25);
      \filldraw[draw=black,fill=black] (0,0) circle (0.07);
    \end{scope} 
  }
}
\tikzset{
  3dpepsdownp/.pic={
    \begin{scope}[canvas is zx plane at y=0]
      \draw (-0.3,0)--(0.3,0);
      \draw (0,-0.25)--(0,0.25);
      \filldraw[draw=black,fill=blue] (0,0) circle (0.07);
    \end{scope} 
  }
}

\newcommand{\cbox}[2]{\vcenter{\hbox{\includegraphics[width=#1em]{#2}}}}

\begin{document}
 
\title{String order parameters for symmetry fractionalization in an enriched toric code}

\author{ Jos\'e Garre-Rubio}
\affiliation{Departamento de An\'alisis Matem\'atico  y Matem\'atica Aplicada, UCM, 28040 Madrid, Spain} 
\affiliation{ICMAT, C/ Nicol\'as Cabrera, Campus de Cantoblanco, 28049 Madrid, Spain}

\author{Mohsin Iqbal}
\affiliation{Max-Planck-Institut f{\"u}r Quantenoptik, Hans-Kopfermann-Stra{\ss}e 1, 85748 Garching, Germany}
\affiliation{Munich Center for Quantum Science and Technology, Schellingstra{\ss}e 4, 80799 M{\"u}nchen, Germany}

\author{David T. Stephen}
\affiliation{Max-Planck-Institut f{\"u}r Quantenoptik, Hans-Kopfermann-Stra{\ss}e 1, 85748 Garching, Germany}
\affiliation{Munich Center for Quantum Science and Technology, Schellingstra{\ss}e 4, 80799 M{\"u}nchen, Germany}

\begin{abstract}

We study a simple model of symmetry-enriched topological order obtained by decorating a toric code model with lower-dimensional symmetry-protected topological states. We show that the symmetry fractionalization in this model can be characterized by string order parameters, and that these signatures are robust under the effects of external fields and interactions, up to the phase transition point. This extends the recent proposal of [New Journal of Physics 21, 113016 (2019)] beyond the setting of fixed-point tensor network states, and solidifies string order parameters as a useful tool to characterize and detect symmetry fractionalization. In addition to this, we observe how the condensation of an anyon that fractionalizes a symmetry forces that symmetry to spontaneously break, and we give a proof of this in the framework of projected entangled pair states. This phenomenon leads to a notable change in the phase diagram of the toric code in parallel magnetic fields

\end{abstract}

\maketitle

\section{Introduction}

Topologically ordered phases have been long studied not just because of their exotic behaviours \cite{Wen90} that are beyond Landau's paradigm, but also for their potential application as error correcting codes \cite{Kitaev03,Dennis02}. 
One of the most salient features of topological phases in 2D is the existence of anyons: quasiparticle excitations that follow neither bosonic nor fermionic statistics \cite{Wilckez82} and recently, direct evidence of their existence has been found \cite{Bartolome20}. 

A paradigmatic example of topological order is the fractional quantum Hall effect \cite{Tsui82, Laughlin83} whose anyons host just a fraction of the electron's charge, {\it i.e.} they fractionalize the charge conservation symmetry. Spin liquids form another prominent example where excitations often carry fractional quantum numbers \cite{Balents2010}. The quantum phases of matter that possess such a non-trivial interplay between symmetries and topological order have been dubbed symmetry enriched topological (SET) phases. 

Bosonic SET phases are well understood: they have been classified \cite{Essin13,Barkeshli14,Mesaros13, Lu16, Barkeshli20}, exactly solvable Hamiltonians for each phase have been constructed \cite{Hermele14, Tarantino16, Heinrich16,Cheng17}, their ground states and anyons have been described in terms of tensor networks \cite{Jiang15,Williamson17,Garre17} and different methods to detect SET phases have been proposed \cite{Huang14, Wang15,Garre19}, to name a few. Notably, most of the aforementioned works focused on renormalization-group (RG) fixed points, but some recent studies of perturbed SET models can be found in \cite{Moon12,Lee16,Nasu17,Wang18}.

In this paper, we study perturbations of a simple fixed-point SET Hamiltonian obtained by decorating a toric code model with lower dimensional symmetry-protected topological orders \cite{Ben-Zion16, TStephen19}. Our main purpose of this study is to understand the extent to which the symmetry fractionalization in this model can be characterized and detected by string order parameters. In \cite{Garre19}, it was shown that string order parameters can be used to detect the symmetry fractionalization in fixed-point projected entangled pair states (PEPS). Here, we wish to extend the range of validity of these order parameters by removing their dependence on tensor network representations, and investigate their behaviour away from fixed-points by driving the model towards phase transitions with various external fields and interactions.

Taking advantage of the simplicity of our fixed-point Hamiltonian and inspired by Ref.~\cite{Garre19}, we define an order parameter which is the expectation value of a certain string operator. We use variational infinite PEPS \cite{vanderstraeten2016gradient, corboz2016variational, haegeman2017diagonalizing} to obtain the ground states of the perturbed Hamiltonian and compute observables in the thermodynamic limit, although we stress that the usage of PEPS is not necessary, and other numerical methods like the density matrix renormalization group \cite{White1992} or variational Monte Carlo \cite{becca2017quantum}, could be used here as well. We find that the order parameter correctly characterizes the symmetry fractionalization within the entire SET phase, with longer string operators providing more accurate results nearer the phase transitions. This shows that string operators also characterize symmetry fractionalization away from fixed-points, and that they can provide a practical tool for numerical detection of SET order.

\begin{figure}[t!]
    \centering
    \includegraphics[width = \linewidth]{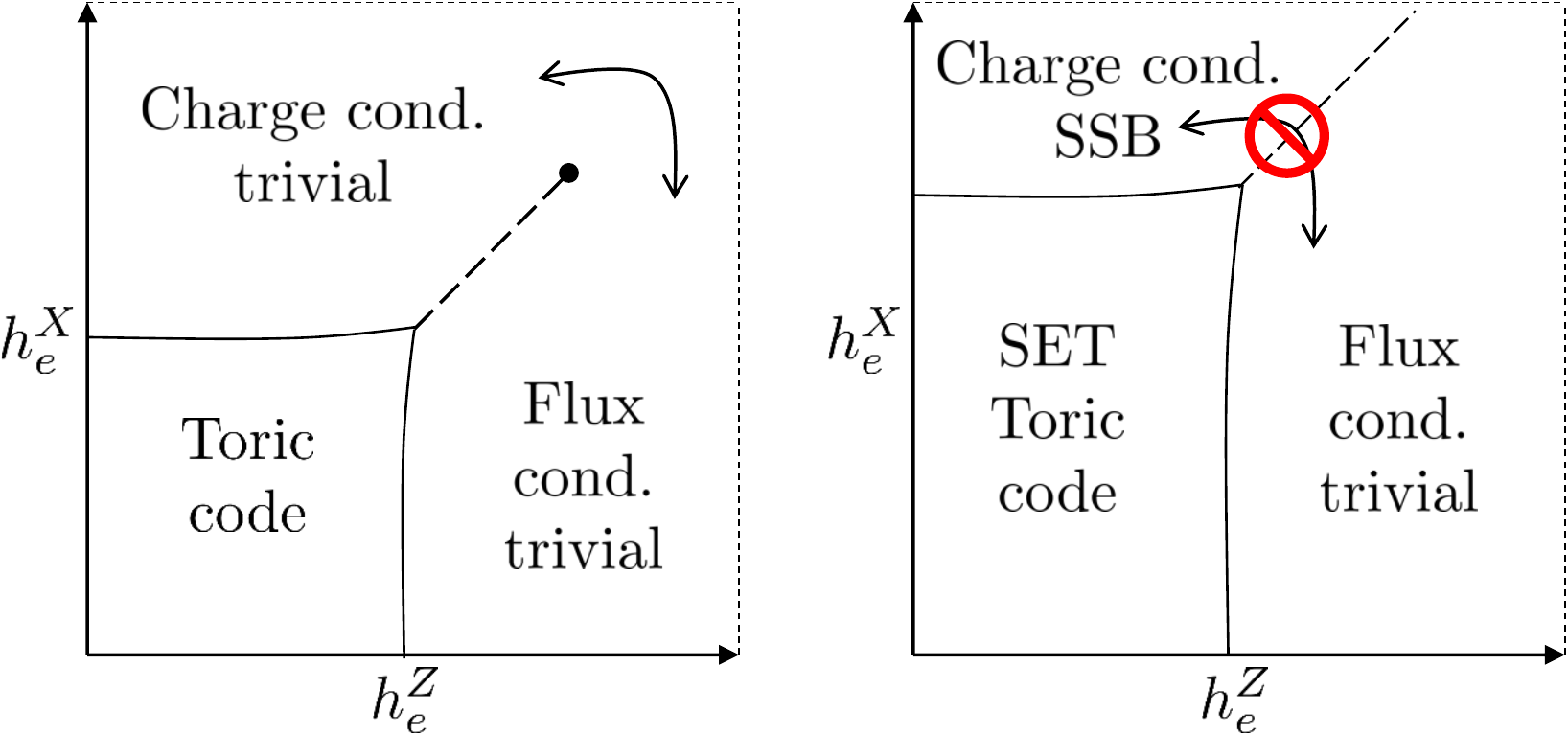}
    \caption{Schematic phase diagram of the toric code (left) and symmetry-enriched toric code (right) in parallel $X$ and $Z$ magnetic fields. Solid (dashed) lines indicate second (first) order phase transitions. The arrows indicate a path connecting the two condensed phases. In the symmetry-enriched case, this path must encounter a phase transitions since the two phases have distinct symmetry breaking patterns.}
    \label{fig:comptc}
\end{figure}

During our investigations, we also observe the phenomenon wherein the condensation of an anyon that fractionalizes the symmetry must result in that symmetry being spontaneously broken. This was proven in Ref.~\cite{Bischoff19} using the framework of $G$-graded tensor categories and it has been identified also in other phase diagrams \cite{Wang18,Sun18}. We also notice that this phenomenon has been studied previously in high energy physics, in particular, it has been proposed to explain the electroweak symmetry breaking \cite{Sannino03}. We provide an alternative proof of this fact in the framework of PEPS. The phenomenon leads to an interesting modification of the well-known 2D phase diagram of the toric code in parallel magnetic fields \cite{trebst2007breakdown,Vidal09A, Tupitsyn10,dusuel2011robustness,Wu12}, see Fig.~\ref{fig:comptc}. We also find that the symmetry is spontaneously broken across phase transitions that cannot be described as condensation transitions, such as that driven by a transverse magnetic field \cite{Vidal09B}, and in the intermediate phase found in a direct interpolation to the toric code phase. This signals that there is yet more to uncover regarding the relation between topological phase transitions and symmetry breaking.

The structure of the manuscript is as follows. In \cref{sec:pertH} we first introduce the target SET Hamiltonian and the string order parameters that we use to capture symmetry fractionalization. Then, we also explain the spontaneous symmetry breaking (SSB) from anyon condensation and we add a proof of this phenomenon in \cref{sec:ssbinT} using PEPS. In \cref{sec:numresults} we show the numerical results of using the SOP on the Hamiltonian interpolation, and we obtain the phase diagrams for different combinations of perturbations applied to the target Hamiltonian. Finally, we summarize our results and discuss their implications towards the characterization of general SET phases in \cref{sec:conclu}.

\section{TC Hamiltonian enriched with cluster state loops} \label{sec:pertH}

In this section, we first define a Hamiltonian which realizes an SET phase and will form the basis of our analysis throughout the paper. Second, we derive the order parameter we use to characterize the SET phases obtained by perturbing the Hamiltonian. Finally, we discuss with the interesting phenomenon of SSB induced by anyon condensation.

\subsection{SET Hamiltonian}

Our SET model can be described as a toric code, whose ground states can be viewed as equal-weight superpositions over closed loop configurations, where these loops are further decorated with 1D symmetry protected topological (SPT) orders, namely cluster states \cite{Raussendorf2003}. Such a model was first proposed in \cite{Ben-Zion16} and also studied in  Refs. \cite{TStephen19,Barkeshli20}. This decoration enriches the toric code with a global $\mathbb{Z}_2\times \mathbb{Z}_2$ symmetry, time-reversal symmetry (TRS), and inversion symmetry which are fractionalized by the charge excitations.

To define the model precisely, let us begin with the toric code with spins on the edges of a honeycomb lattice. The Hamiltonian is
\begin{equation}
H_{TC} = -\sum_{v\in V} A_v - \sum_{f\in F} B_f,
\end{equation}
where $V$ ($F$) denotes the set of all vertices (faces) in the lattice and the terms in the Hamiltonian are
\begin{equation}
A_v=\prod_{e\ni v}Z_e\;  \text{ and } \; B_f=\prod_{e\in f}X_e \ ,  
\end{equation}
where $X$ and $Z$ are the spin-1/2 Pauli operators, $e \ni v$ denotes all edges terminating on $v$ and $e\in f$ denotes all edges surrounding $f$. Let $\mathcal{C}$ be a subset of edges that form closed loops on the lattice. We denote by $|\mathcal{C}\rangle$ the state where all edges in $\mathcal{C}$ are in the state $|1\rangle$ and the rest are in $|0\rangle$. 
Then, the ground states of $H_{TC}$ can be written as an equal-weight superposition of loop configurations: 
\begin{equation}
|TC\rangle = \frac{1}{\mathcal{N}}\sum_{\mathcal{C}} |\mathcal{C}\rangle \ ,
\end{equation}
for some normalization factor $\mathcal{N}$. Here, we are assuming trivial topology such that $|TC\rangle$ is the unique ground state of $H_{TC}$.

Now, we introduce new spins on the vertices of the lattice and initialize them in the state $|+\rangle^{|V|} = \bigotimes_{v \in V}\frac{1}{\sqrt{2}}(|0\rangle_v + |1\rangle_v)$. We couple these to the edge spins using a unitary circuit $U_{CCZ}$ which is constructed using $CCZ$ operators acting on every triplet consisting of an edge $e$ and its two vertices $v_e^{+,-}$, where $CCZ = |0\rangle\langle 0|_e \otimes \mathbb{I}_{v_e^+ v_e^-} + |1\rangle\langle 1|_e\otimes CZ_{v_e^+ v_e^-}$ and $CZ_{v_e^+ v_e^-} = |0\rangle\langle 0|_{v_e^+}\otimes \mathbb{I}_{v_e^-} + |1\rangle\langle 1|_{v_e^+}\otimes Z_{v_e^-}$. We denote the resulting state as
\begin{equation} 
|SET\rangle = U_{CCZ} \left(|TC\rangle \otimes |+\rangle^{|V|} \right).
\end{equation}

This circuit acts as $CZ$ along the vertices of a loop and it acts trivially away from them. Acting on initial $|+\rangle$ states with $CZ$'s between nearest neighbours creates the 1D cluster state. Thus, the vertices along loops form cluster states, which are an example of 1D SPT orders \cite{Son2012}. Decorating loops with 1D SPT orders is a well-known way of creating SET orders \cite{Li14,Huang14,Barkeshli14,Ben-Zion16}. 

\begin{figure}
    \centering
    \includegraphics[width=0.7\linewidth]{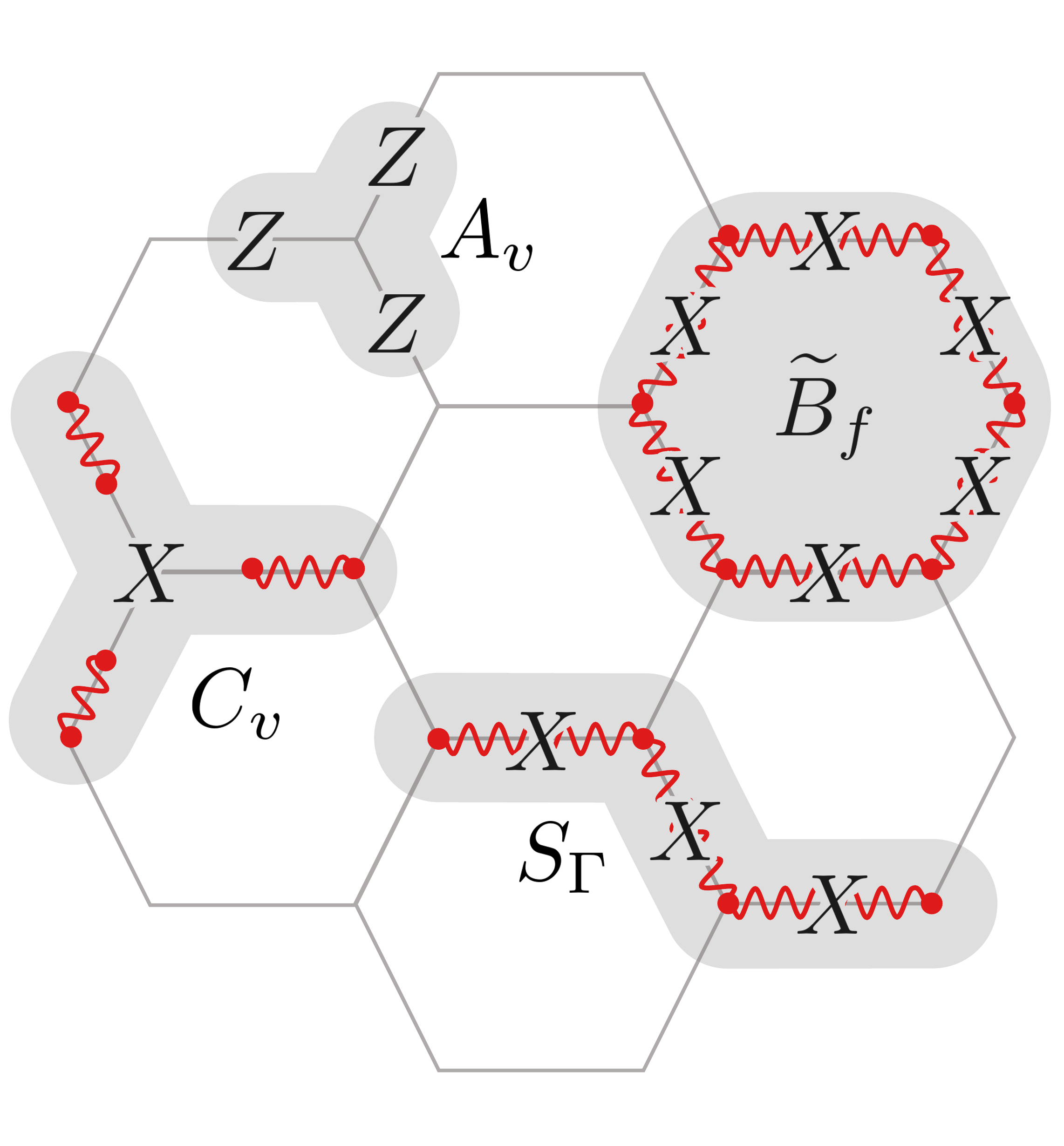}
    \caption{Hamiltonian terms in $H_{SET}$ (Eq.~(\ref{eq:H_{SET}})) and an example of a string operator $S_\Gamma$ that creates pairs of charge excitations at the endpoints (Eq.~\ref{eq:sgamma}). A wavy line connecting two sites marked with circles represents a $CZ$ operation on these sites.}
    \label{fig:ham}
\end{figure}

To obtain the Hamiltonian for which $|SET\rangle$ is the ground state, we can simply conjugate the initial uncoupled Hamiltonian
\begin{equation} \label{eq:H_triv}
\tilde{H}_{TC} = H_{TC} - \sum_{v\in V} X_v,
\end{equation} by $U_{CCZ}$ to obtain
\begin{equation} \label{eq:H_{SET}}
H_{SET}  = -\sum_{v\in V} A_v - \sum_{f\in F} \widetilde{B}_f - \sum_{v\in V} C_v \frac{1+A_v}{2}
\end{equation}
where $A_v$ is unchanged from before, and $\widetilde{B}_f$ is a modified version of $B_f$ decorated by additional $CZ$ operators, see Fig.~\ref{fig:ham}. $C_v = U_{CCZ} X_v U_{CCZ}^\dagger$ is also defined pictorially in Fig.~\ref{fig:ham}. We have additionally modified $C_v$ by adding a projector $\frac{1+A_v}{2}$ onto the closed-loop subspace. 
This only affects the energy of some excitations, and serves the purpose of making $H_{SET}$ commute with a $\mathbb{Z}_2\times\mathbb{Z}_2$ symmetry group generated by $X_A$ and $X_B$, which are defined to flip every spin on all vertices of the $A$ and $B$ sublattice, respectively. Since the Hamiltonian is real, it also has time-reversal symmetry generated by $\mathcal{K} X_A X_B$, where $\mathcal{K}$ denotes complex conjugation. Moreover, the Hamiltonian is invariant under bond-centered inversion of the lattice.

\subsection{Symmetry fractionalization in $H_{SET}$}

$H_{SET}$ represents a non-trivial SET order in the presence of either the $\mathbb{Z}_2\times\mathbb{Z}_2$ symmetry or the time-reversal symmetry. To see why, consider the following string operator,
\begin{equation} \label{eq:sgamma}
S_\Gamma  = \prod_{e\in \Gamma} X_e CZ_{v^+_e v^-_e} 
\end{equation}
where $\Gamma$ is a path of edges and $v_e^{+,-}$ are  the two vertices attached to $e$, see Fig.~\ref{fig:ham}. This operator creates charge excitations at the endpoint vertices $v_i$ and $v_f$ of $\Gamma$ corresponding to violations of $A_v$. Because $A_v = -1$ at these vertices, the Hamiltonian term $C_v$ is disabled by the projector $\frac{1+A_v}{2}$. Therefore, we can dress the endpoint vertices of $S_\Gamma$ with $Z$ operators without changing the energy of the excitation,
\begin{equation} \label{eq:chargeop}
S_\Gamma(\alpha,\beta) = Z_{v_i}^\alpha Z_{v_f}^\beta S_\Gamma.
\end{equation}
This gives a four-fold degenerate subspace of states associated to each $\Gamma$ spanned by the states $|\alpha,\beta\rangle :=S_\Gamma(\alpha,\beta)|SET \rangle$. Within this subspace, the $\mathbb{Z}_2\times\mathbb{Z}_2$ symmetry acts projectively on each charge as
\begin{align}
&X_{A}|\alpha,\beta\rangle=|\alpha \oplus 1, \beta\oplus 1\rangle. \nonumber \\
&X_{B}|\alpha,\beta\rangle=(-1)^\alpha(-1)^\beta|\alpha,\beta \rangle,
\end{align}
where we have assumed that $\alpha$ and $\beta$ are both odd vertices, a similar result holds in the general case (see \cite{Ben-Zion16}). If we let $V_{(1,0)}$ ($V_{(0,1)}$) denote the local action of $X_A$ ($X_B$) on a single charge, we have $V_{(1,0)} = X$, $V_{(0,1)} = Z$ and $V_{(1,1)}=XZ$, where we have labelled $\mathbb{Z}_2\times \mathbb{Z}_2 = \{(a,b): a,b=0,1\}$. Since $X$ and $Z$ anticommute, the charge carries a projective representation of $\mathbb{Z}_2\times\mathbb{Z}_2$, which demonstrates the fractionalization. Concretely, if we write $V_qV_k=\omega(q,k)V_{qk}$, where  $q,k\in \mathbb{Z}_2\times \mathbb{Z}_2 $, the symmetry fractionalization (SF) pattern is given by
\begin{equation}\label{SFenrtc} \omega(q,k)=
  \left( {\begin{array}{cccc}
   +1 & +1 & +1&+1 \\
 +1 & +1 &  -1 &- 1 \\
 +1 & +1 & +1 & +1 \\
+1 & +1 &  -1 & -1 
  \end{array} } \right)_{q,k}.
\end{equation}
The local action of $\mathcal{K} X_A X_B$ is $\mathcal{T}=XZ$ so that $\mathcal{T}^2 = -1$. Then, the charge also fractionalizes time-reversal symmetry. Finally, since $X_A$ and $X_B$ anticommute near a charge, and inversion swaps $A\leftrightarrow B$, it follows that inversion anticommutes with $X_AX_B$ near a charge. Conversely, it is easy to see that the SF pattern of the charge for $\tilde{H}_{TC}$ of \eqref{eq:H_triv} is $\omega(q,k)=+1$ for all $q,k\in \mathbb{Z}_2\times\mathbb{Z}_2$, so that it is trivial. In Appendix \ref{sec:TNGS} we construct a PEPS representation of $|SET\rangle$ which provides another viewpoint on the SF pattern.

We note that the flux excitations are created by a string of $Z$ operators corresponding to a path on the dual lattice, and they have no symmetry fractionalization. The dyon, which is a composite of flux and charge, fractionalizes in the same way as the charge.

\

Since we have used decoration by 1D SPT phases to construct our 2D SET model, some comments on their classifications are in order. For a global on-site symmetry group, $Q$, 1D SPT phases are classified by the group $H^2(Q,U(1))$. When $Q= \mathbb{Z}_2\times \mathbb{Z}_2$, $H^2(Q,U(1))= \mathbb{Z}_2$ which means that there are only two phases: the trivial one and the non-trivial SPT phase (the one of the cluster state, {\it i.e.} the Haldane phase). In the case of SET phases, the classification of SF patterns is given by  $H^2(Q,G)$, where $G$ is an abelian group that depends on the topological order. In the present case, a toric code enriched with a $\mathbb{Z}_2\times \mathbb{Z}_2$ symmetry that fractionalizes the charge and not the flux (without anyon permutation) could host four inequivalent SF patterns.

By simple counting, this implies that not all SET phases can be constructed by decorating topological orders with 1D SPT phases as we have done here. This fact affects how to detect SET phases since it implies that 1D SPT order parameters \cite{Pollmann12,Haegeman12}, or their embedding into 2D \cite{Huang14}, cannot give a complete characterization of the SET phases (in the case of toric code topological order with $G=\mathbb{Z}_2\times\mathbb{Z}_2$ symmetry, they can say that the SET phase is non-trivial but they don't discriminate between the different non-trivial ones). That is the reason why in the next section, we study a generalization of the order parameters first defined in \cite{Garre19} which can completely characterize SET phases. We remark that, for time reversal symmetry, where there are only two phases, the SF pattern of the charge can be characterized using 1D SPT order parameters \cite{Huang14}. 

\subsection{Order parameters for SET phases}

To measure the effect of a perturbation on the SET phase, we need to use an order parameter which detects the SF pattern of the anyons. Such an order parameter was proposed in Ref.~\cite{Garre19}. Therein, the authors proposed a set of string order parameters (SOPs) $\mathcal{O}^{[q]}$ index by elements $q$ of the on-site symmetry group $Q$ ($=\mathbb{Z}_2\times\mathbb{Z}_2$ here), that, for topological PEPS with zero correlation length, reveal the lower diagonal $\{\omega(q,q); 0 \neq q\in Q\}$ of the SF pattern, which is sufficient to completely characterize the phase. The value of $\omega(q,q)$ characterizes the action of $q$ applied twice on the charge. Since in our case $q \in \mathbb{Z}_2\times\mathbb{Z}_2 $, {\it i.e.} the group is order two, a value $\omega(q,q)\neq 1$ implies a projective action on the charge, see \eqref{SFenrtc}. Here, we modify the SOPs in order to extend their range of applicability beyond fixed-point tensor network states. 

To begin, let us summarize the SOPs of Ref.~\cite{Garre19}. Concretely, the goal of the order parameter is to capture the projective action of the symmetry on a charge. This action is equal to the effect of the braiding between the charge and some other anyon \cite{Barkeshli14}. The key idea to exploit this connection is to use how braiding, and thereby the projective action is detected: via an overlap of the anyon affected by the braiding with a non-affected one. The SOP evaluates the overlap of two charges placed at the ends of a tensor product of symmetry operators together with \swap operators that permute sites. The symmetry operators and the \swap are meant to reproduce the effect of acting with the symmetry twice over an isolated charge, to capture the value of $\omega(q,q)$.

One limitation of the SOPs described above is that they rely on a tensor network representation of the ground state in order to create an excited state with charges. Even when such a representation is available, the order parameters of Ref.~\cite{Garre19} require imposing an additional topological structure on the tensor network, which makes variational optimization of the tensors more costly. As such, it is desirable to modify the SOPs in such a way that they are not reliant on the structure provided by topological tensor networks. For this purpose, we now introduce a new set of SOPs that are designed to detect the SF pattern of $H_{SET}$, and have no reliance on tensor networks. To insert charges into the ground state, we use the string operator $S_\Gamma$ (Eq. \ref{eq:sgamma}) of the fixed point Hamiltonian. When $H_{SET}$ is perturbed, the state created by acting with $S_\Gamma$ on the ground state will no longer be an eigenstate in general, but we nonetheless expect it to have finite overlap with the corresponding eigenstate having charges at the endpoints of $\Gamma$. Since  the SOPs are defined by a ratio of two expectations values, we do not expect this to be an issue. We remark that for detecting the SF pattern, we assume the knowledge of the topological order, in particular the explicit form of the anyon creation operators (at some point of the phase diagram).

The SOPs we use are given in terms of the expectation value of some string operators $\Lambda^{[q]}$, that depend  on the elements $q\in  \mathbb{Z}_2 \times \mathbb{Z}_2 $ of the on-site symmetry group. To define the string operator let us consider a line of $5$ vertices with their $4$ inside edges on the hexagonal lattice and define
\begin{align}
\label{sophex}
\Lambda^{[a,b]}= &
 \;  S^\dagger_{\Gamma = 1,\cdots,6}  [ X^a_1 X^b_2 X^a_3   X^b_4]
\times \notag  \\
& [ \swap_{e_1,e_3} \swap_{e_2,e_4} \swap_{1,3}\swap_{2,4}] 
 S_{{\Gamma}=5,6}  \notag \\
   =& \; \left( \prod_{i=1}^4 CZ_{i,i+1} X_{e_i} \right) (X^a_1 X^b_2 X^a_3   X^b_4)  \times \notag  \\  & \;  \swap_{e_1,e_3} \swap_{e_2,e_4} \swap_{1,3}\swap_{2,4}. 
\end{align}
This latter equality can be written pictorially as:
\begin{equation}
\cbox{20}{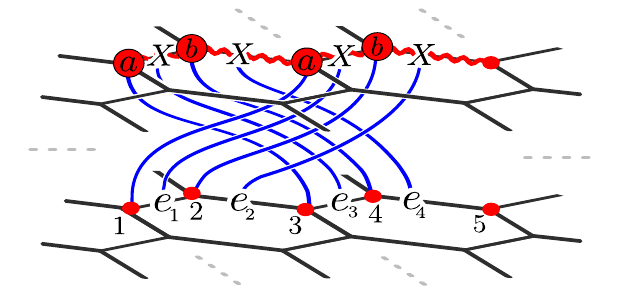},
\end{equation}
where circles labelled $a$ and $b$ denote symmetry operators $X^a$ and $X^b$, wavy red lines depict $CZ$ gates between neighbouring vertices, and blue lines depict the action of the \swap operators. Using $\Lambda^{[a,b]}$ we define the following SOPs:

\begin{equation} \label{eq:newop}
\mathcal{O}^{[a,b]} = \frac{ \langle \Lambda^{[a,b]} \rangle}{\langle \Lambda^{[0,0]} \rangle} 
,
\end{equation}
where the expectation value $\langle \; \cdot \; \rangle $ is taken with respect to the ground state of the perturbed Hamiltonian. Our SOP is derived using the same physical principles as in Ref.~\cite{Garre19}, but we remark that it is not the direct translation of the SOP from Ref.~\cite{Garre19} into a physical operator, since that would result in a fattened string operator acting on a line of plaquettes. Rather, we have chosen to slim the order parameter to act along a single line, easing numerical calculation.

For systems with a non-zero correlation length, {\it i.e.} away from the fixed points, we can define a family of SOPs of length $\ell$ which we expect will better capture the SF pattern with increasing $\ell$, as in the case of 1D SOPs \cite{Pollmann12,Haegeman12}. The corresponding operator acts on $4\ell + 1$ vertex spins,
\begin{align}
    \Lambda_\ell^{[a,b]} = &
    \prod_{j=1}^{4\ell} CZ_{j,j+1} X_{e_j}
    \prod_{i=1}^\ell  \Big[  X^a_{2i-1} X^b_{2i} X^a_{2i+2\ell-1} X^b_{2i+2\ell} \nonumber \\
    &  \times
    \swap_{2i-1,2i+2\ell-1} \swap_{2i,2i+2\ell}  \nonumber \\
    &  \times
     \swap_{e_{2i-1},e_{2i+2\ell-1}} \swap_{e_{2i},e_{2i+2\ell}} \Big ]
\end{align}
so that the SOP of length $\ell$ is defined as follows:

\begin{equation} \label{eq:opgenerl}
    \mathcal{{O}}_\ell^{[a,b]} = \frac{ \langle \Lambda_\ell^{[a,b]} \rangle }{ \langle \Lambda_\ell^{[0,0]} \rangle }.
\end{equation}
This SOP is expected to better capture the SF pattern with increasing $\ell$, namely,
\begin{equation}\label{expbesop}
     \lim_{\ell \to \infty} \mathcal{{O}}_\ell^{[a,b]} = \omega((a,b),(a,b)) \ .
\end{equation}
We will focus mainly on $\mathcal{O}_\ell^{[1,1]}$ since it is the one that distinguishes the SF patterns of $\tilde{H}_{TC}$ and $H_{SET}$ (because $\omega((0,1),(0,1)) =\omega((1,0),(1,0))=1$ for both systems).

In Appendix \ref{app:TCPEPS}, we give a first test of these order parameters by applying symmetry-preserving deformations to a certain fixed-point PEPS tensor which drive across a phase transition to a topologically trivial phase. We find that, as the length $\ell$ is increased, the value of the SOP converges to the value corresponding to the fixed-point tensor as expected. In contrast, when the deformation explicitly breaks the symmetry, the order parameters show no clear signature of convergence.

We warn that these SOPs are defined for SET phases, that is, topological phases with an unbroken symmetry. Because of this, the meaning of the value of \eqref{eq:opgenerl} after the phase transition point is not clear if the perturbation drives the model to a non-SET phase, such as one where the symmetry is spontaneously broken or the topological order is trivial. For example, when inversion symmetry is broken, the order parameter can depend on whether the endpoints lie on the $A$ or $B$ sublattice. Furthermore, we will find that the value becomes undefined in trivial phases where the charge excitation is confined since in that case, the expectation value of $S_\Gamma$ goes to zero (and then $\langle \Lambda_\ell^{[0,0]} \rangle$ goes to zero as well).

Finally, we remark that a similar procedure can be carried out for the other PEPS order parameters of Ref.~\cite{Garre19}. In general, to detect other SF patterns, it is necessary to change which sites are permuted by the \swap depending on the topological order and the symmetry group \cite{Garre19}. The use of the charge in this explanation is motivated by the SF pattern of $H_{SET}$ but it could be applied to any anyon that fractionalizes a global on-site symmetry.

\subsection{Spontaneous symmetry breaking from anyon condensation} \label{sec:ssb}

In this section, we explain one interesting phenomenon that we will observe in the phase diagrams obtained by perturbing $H_{SET}$. We find that the ground state subspace exhibits spontaneous symmetry breaking (SSB) when an anyon that fractionalizes the symmetry (the charge) is condensed. We note that this has been proven in full generality, using the language of $G$-graded tensor categories, to be a necessary outcome of condensing a fractionalized anyon \cite{Bischoff19}. In \cref{sec:ssbinT} we give an alternative proof of this based on PEPS. Our proof aims to reach a broader audience using a simpler mathematical formalism, at the cost of some generality.

To understand the physics behind this phenomenon, let us make the following cartoon picture. The starting point is some initial topological phase, invariant under some global symmetry, which is undergoing an anyon condensation process. We denote by $b$ the anyon that is condensing (we assume there is only one for the sake of simplicity) and by $1$ the vacuum of the initial topological phase. The vacuum of the final phase, $\varphi$, is an anyon condensate of $b$ together with $1$, which we can write as $\varphi = 1 + b$. If $b$ transforms non-trivially under the symmetry, this affects how the composition $1+b$, and therefore $\varphi$, behaves under the action of the symmetry. 

In the case where $b$ fractionalizes the symmetry and the symmetry is preserved, this would imply that either $\varphi$ transforms projectively, or in  an undefined way since it is a superposition of $b$ and $1$. However, a vacuum that transforms non-trivially is not a valid theory, so the symmetry has to break spontaneously in order to act trivially on the vacuum.

For the case where $b$ is permuted to another anyon $c$ of the theory, $\varphi$ would be also identified with $1+c$. But $c$, in general, will be of another nature of $b$ in the anyon condensation process, that is, it can be a confined anyon. Therefore, since by definition no confined anyon can condense, the symmetry has to break spontaneously to consistently act on the vacuum. This leads to the conclusion that the set of condensed anyons must be closed under the permutation action of the symmetry.

\section{Phase diagram of $H_{SET}$} \label{sec:numresults}

In this section, we first study a Hamiltonian interpolation between $H_{SET}$ and $\tilde{H}_{TC}$ to understand the behaviour of the SOPs of Eq.~\eqref{eq:opgenerl} in the different SET phases and how they depend on $\ell$. Afterward, we introduce perturbations to $H_{SET}$ that commute with the symmetry (or part of it) and we study the resulting phase diagrams numerically. We focus on the behaviour of the SOPs as well as the possibility of SSB driven by anyon condensation. For the latter, we employ local order parameters $\langle Z_{v_A}\rangle$, $\langle Z_{v_B}\rangle$, and $\langle X_{v_A} - X_{v_B}\rangle$ which anticommute (only) with $X_A$, $X_B$, and inversion, respectively, where $v_{A/B}$ denotes a vertex on the $A/B$ sublattice. We also measure local magnetizations on the edge degrees of freedom to help understand the structure of the perturbed ground states.

The perturbations that entirely commute with the symmetry are magnetic fields ${\bm{\sigma}}_e=\left( X_e,Y_e,Z_e\right)$ on edges that are meant to break the topological order in different ways. We also apply a $ZZ$ Ising interaction on the vertices of $H_{SET}$ that changes the symmetry fractionalization pattern, since it breaks explicitly the $\mathbb{Z}_2\times\mathbb{Z}_2$ symmetry down to a $\mathbb{Z}_2$ subgroup generated by $X_AX_B$, but preserves the topological order. In the next subsections, we examine the result of applying each perturbation separately. We also consider the combination of $X$ and $Z$ fields and examine their 2D phase diagram.

We use the iPEPS algorithm to approximate the ground states of the perturbed Hamiltonians \cite{vanderstraeten2016gradient, corboz2016variational, haegeman2017diagonalizing}.
The unit tensor that we use for the PEPS description consists of 5 (i.e., 3 edge and 2 vertex) spins on the hexagonal lattice, as depicted in Appendix \ref{sec:TNGS}. It is important to note that we choose a local tensor that contains vertices $v_A$ and $v_B$ to allow for the spontaneous breaking of inversion symmetry. The variational manifold is characterized by $3^4 \times 2^5 = 2592$ complex parameters, as we use the bond dimension $D=3$ for iPEPS simulations (note that the ground state of the fixed-point Hamiltonian admits an exact representation with $D=3$, as shown in Appendix \ref{sec:TNGS}). We remark that even though the bond dimension is small, the number of variational parameters makes the tensor network ansatz rich enough to capture the qualitative features of the phase diagram. The optimization is comprised of iteratively updating the local tensor by Broyden–Fletcher–Goldfarb–Shanno (BFGS) algorithm \cite{broyden1970convergence,fletcher1970new,goldfarb1970family,shanno1970conditioning}. The energy gradient with respect to the local tensor has been computed by contracting tensor networks on a cylindrical geometry that is infinite along one direction and we choose a gradient norm of the order $10^{-6}$ as a stopping criterion for the optimization algorithm.  Finally, the reported expectation values of the local observables and SOPs have been computed on the infinite plane by the boundary MPS method \cite{haegeman2017diagonalizing}, where we choose the bond dimension of the environment tensors to be of the order $5D^2$.

\subsection{Hamiltonian interpolation}
\label{sec:Hinter}
\begin{figure}[h!]
	\includegraphics[width=\linewidth]{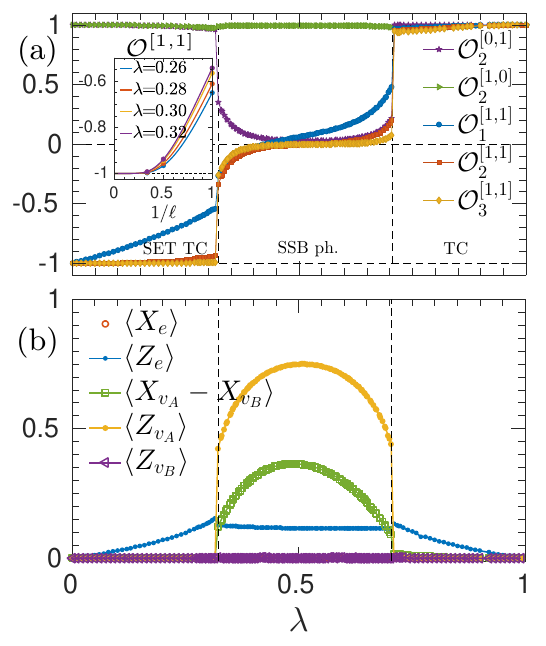}
	\caption{(a) SOPs and (b) magnetizations per site across two phase transitions governed by the Hamiltonian interpolation. The asymmetry between $\mathcal{O}^{[0,1]}_3$ and $\mathcal{O}^{[1,0]}_3$ in the intermediate phase is due to the spontaneous breaking of inversion symmetry for the chosen ground state that is shown. The inset in (a) shows the convergence of $\mathcal{O}^{[1,1]}$ to $-1$ near the phase transition in the SET TC phase. Solid lines and circles indicate the fit [$a\exp(-b\ell)-1$] and the computed data points, respectively.}
	\label{fig:interpolation}
\end{figure}

In this section we check the behaviour of the SOPs on the resulting phase diagram of the following Hamiltonian interpolation:
\begin{equation}
H(\lambda)= \lambda \tilde{H}_{TC} +(1-\lambda)H_{SET},
\end{equation}
such that $H(0)=H_{SET}$ and $H(1)=\tilde{H}_{TC}$. This path therefore interpolates between an SET phase with a non-trivial SF pattern and one with trivial SF. In Fig.\ref{fig:interpolation} we show the numerical results for the evaluation of the order parameters and the different local magnetizations. 

We see in Fig.\ref{fig:interpolation}(a) that the order parameter correctly characterizes the SF patterns of the two ends of the interpolation. In particular, $\mathcal{O}^{[1,1]}_\ell$ is exactly $-1$ at the SET fixed-point ($\lambda=0$) and $+1$ at the toric code point ($\lambda=1$). Surprisingly it goes to zero in the middle part, pointing towards an intermediate phase between the two SET phases. The other order parameters $\mathcal{O}^{[0,1]}_1$ and $\mathcal{O}^{[1,0]}_1$ are equal to $1$ at each endpoint, such that Eq.\eqref{expbesop} is exactly satisfied at the two fixed-points.

Away from the fixed-points, we can see that $\mathcal{O}^{[1,1]}_\ell$ gets sharper for increasing $\ell$ and approaches the fixed-point value of each region  (see blue, red and yellow lines of Fig.\ref{fig:interpolation}(a) corresponding to $\ell=1,2,3$). From the inset of Fig.~\ref{fig:interpolation}(a), we see that the SOP appears to converge towards the fixed-point value in the SET phase exponentially with string length, although our computational resources do not allow us to go beyond $\ell = 3$, making it difficult to make any rigorous claims. Altogether, these results affirm Eq.\eqref{eq:opgenerl} as a reliable order parameter to probe SET phases.

The nature of the intermediate phase can be seen clearer by the behaviour of the magnetizations in Fig. \ref{fig:interpolation}(b). This intermediate phase spontaneously breaks all of the symmetries, as indicated by non-zero values of the SSB order parameters. We note that, by nature of our numerical technique (that involves initialization of tensors for the variational optimization with an optimal ground state tensor of a nearby point), we see only one of the SSB ``branches'' corresponding to one of SSB ground states.
For other ground states, $\langle Z_{v_A}\rangle$ is zero while $\langle Z_{v_B} \rangle$ is non-zero. We have also checked that it is possible to construct ground states in all the branches if we randomly initialize the variational optimization at every point in the intermediate phase.
We defer a closer examination of the SSB pattern to Sec.~\ref{subsec:SSBtc}, since it is the same in both cases.

We point out that the interpolation of $(\mathbb{Z}_2)^3$ SPT phases studied in \cite{Dupont20} can be connected to ours by gauging a $\mathbb{Z}_2$ subgroup of the $(\mathbb{Z}_2)^3$ symmetry \cite{TStephen19}. However, this is not an exact duality due to the fact that we additionally added the projection $(1+A_v)/2$. We leave for future work the details and implications of this connection.

\subsection{$X_e-Z_e$ fields: transitions to trivial phases}

We now present our findings regarding the phase diagram of $H_{SET}$ in the presence of parallel $X$- and $Z$-fields on the edges:
\begin{equation}
	H_{SET} - h^{X}_e\sum_{e}Z_e - h^{Z}_e\sum_{e}X_e.
\end{equation}
While the phase diagram exhibits some similarities to the well-known phase diagram of regular toric code \cite{trebst2007breakdown,Vidal09A, Tupitsyn10,dusuel2011robustness,Wu12}, it also has some novel features.  In analogy to the regular toric code, the $X$- and $Z$-fields' action drives the condensation of charge and flux excitations, respectively.  However, as discussed in Sec.~\ref{sec:ssb}, the presence of an anyon that fractionalizes the symmetry leads to SSB after charge condensation, which leads to a different phase diagram.

\begin{figure}[t!]
		\includegraphics[width = \linewidth]{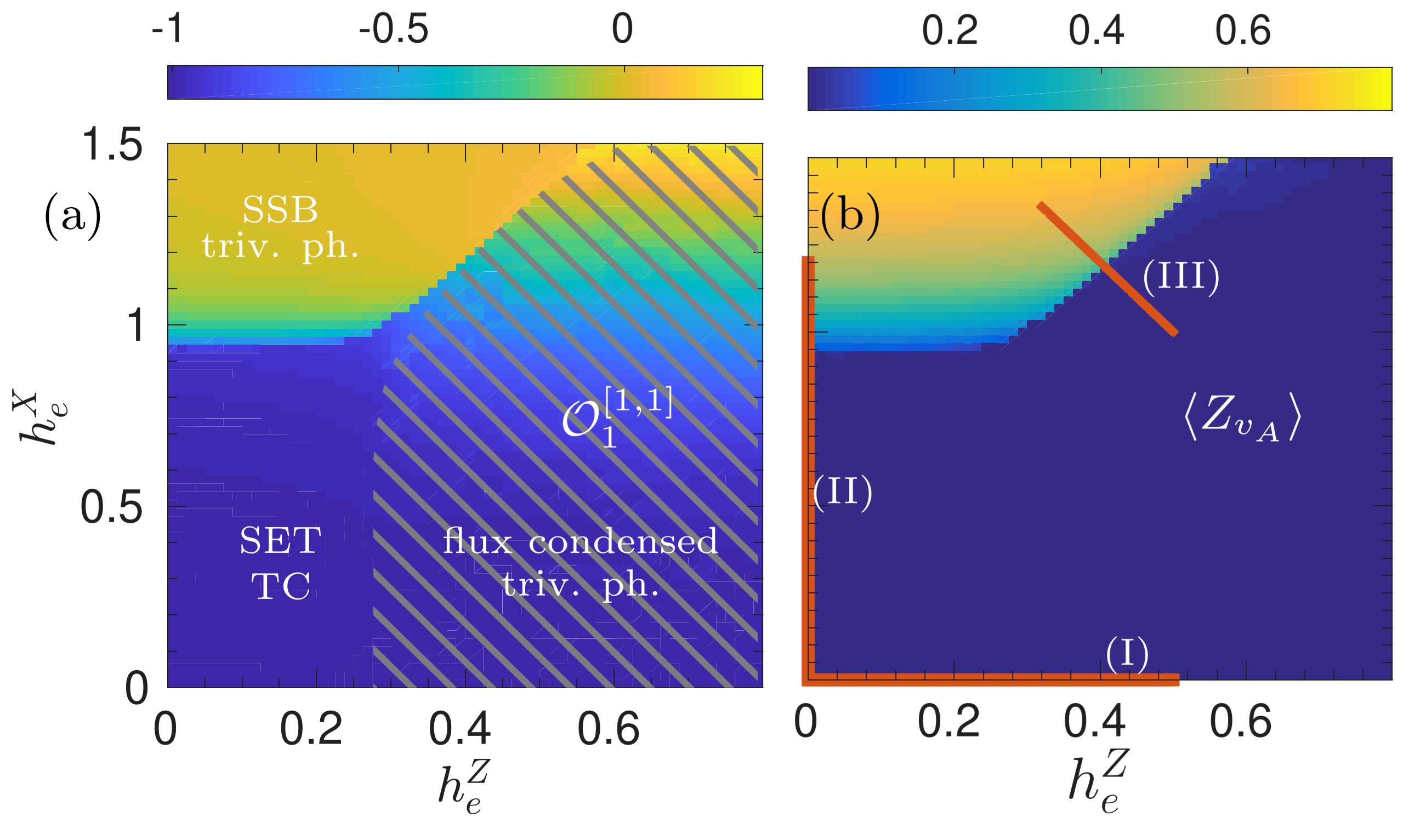}
	\caption{ Phase diagram of $H_{SET}$ with $X$- and $Z$-fields showing $\mathcal{O}_1^{[1,1]}$ and $\langle Z_{v_A}\rangle$ on the left and right respectively. On the right we indicate the three lines that we study in detail. The crossed-out region is where the charge is confined, making $\mathcal{O}_1^{[1,1]}$ ill-defined.}
	\label{fig:phdiag_xz}
\end{figure}

Fig.~\ref{fig:phdiag_xz} shows the 2D phase diagram in terms of the order parameter   $\mathcal{O}_1^{[1,1]}$ and the vertex $Z$-magnetization $\langle Z_{v_A}  \rangle$ ($\langle Z_{v_B} \rangle$ is 0 through the phase diagram for the particular symmetry-breaking ground state we obtain).  We see that the string order parameter is able to detect the entire SET phase in a broad 2D range of the phase diagram (the crossed-out region indicates where the charge is confined, such that $\mathcal{O}_1^{[1,1]}$ becomes ill-defined). The $Z$-magnetization detects the transition to the charge-condensed trivial phase, and in that phase indicates that the symmetry is spontaneously broken. This supports the general claim given in \cref{sec:ssb} that the condensation of a fractionalized excitation (the charge) must be accompanied by SSB. As in the usual toric code phase diagram (with no symmetry fractionalization), there is a line of phase transitions between the two trivial condensed phases. In the toric code case, this line ends at a finite point, meaning that the two condensed phases are in fact the same trivial phase. Here due to the non-trivial SF, we expect that the line cannot end at a finite point since the two topologically trivial phases differ in their symmetry-breaking pattern.

In the following, we further study the lines labeled as (I), (II), and (III) in Fig.~\ref{fig:phdiag_xz}(b) to access the universal features of the three transition lines in the phase diagram.

\subsubsection*{Line(I): flux condensation via $Z_{e}$-field }

\begin{figure*}[t!]
	\includegraphics[scale=2.4]{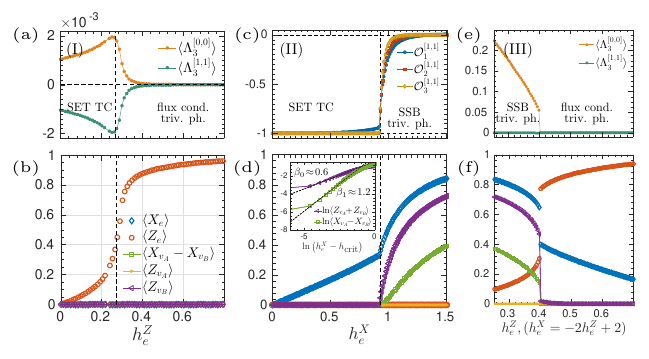}
	\caption{(a),(b): Line(I) Phase transition from SET to the flux condensed trivial phase. (c),(d): Line(II) Transition of the SET to a charge condensed phase with SSB. (e),(f): Line(III) Phase transition between the charge condensed (SSB) and the flux condensed (symmetric) phase.}
	\label{fig:L123b}
\end{figure*}

Here we study the one-parameter Hamiltonian:
\begin{equation} \label{eq:pertz}
H_{SET} - h^{Z}_e\sum_{e}Z_e.
\end{equation}
The $Z_e$-field commutes with the circuit $U_{CCZ}$ that was used to construct $H_{SET}$. Therefore, the same physics as the regular toric code in a $Z$ field primarily governs the resulting phase transition \cite{trebst2007breakdown,Vidal09A, Tupitsyn10,dusuel2011robustness,Wu12}. Indeed, our numerics are compatible with the transition point of the regular toric code in $Z$-field, which is dual to the transverse field Ising model on a triangular lattice \cite{Blote2002}. The $Z_e$-field penalizes configurations with longer loops so that it condenses the flux. As a consequence of the breakdown of the  topological order due to flux condensation, we observe an increase in the expectation value of the $Z_e$-field on the edges (Fig. \ref{fig:L123b}(b)). All SSB order parameters are 0 and we find that there is one symmetric ground state, {\it i.e.} no symmetry breaking. This is expected, since the flux does not fractionalize any symmetry.

From Fig.\ref{fig:L123b}(a) we see that $\mathcal{O}_3^{[1,1]} = -1$ exactly for any value of $h^Z_e$, so the SOP works as expected in the SET phase. However, the value after the phase transition is ill-defined since, dual to the condensation of the flux, the phase transition also confines the charge. This effect can be measured by the confinement fraction of the charge \cite{Duivenvoorden17,Iqbal18} which is given by $\langle \Lambda^{[0,0]} \rangle$. As shown in Fig.\ref{fig:L123b}(a),  $\langle \Lambda^{[0,0]} \rangle$ goes to zero after the phase transition, indicating the charge confinement and invalidating the use of $\mathcal{O}^{[a,b]}$ after this point.

\begin{table}[t!]
   \caption{\label{tab:ssb}Values of SSB order parameters for the different short-range correlated ground states in the charge condensed SSB phase (upper half) and ${\rm Ising}_A\otimes {\rm Ising}_B$ model (lower half). A marking of $\pm$ indicates an arbitrary positive or negative value.}
\setlength\tabcolsep{0pt}

            \begin{tabular*}{0.9\linewidth}{@{\extracolsep{\fill}}c c c c}
            \hline\hline
        GS & $\langle Z_{v_A} \rangle$ & $\langle Z_{v_B} \rangle$ & $\langle X_{v_A} - X_{v_B} \rangle$  \\
        \hline
        1 & $+$&  0 & $+$ \\
        2 & $-$&  0 & $-$ \\
        3 & 0 &  $+$& $-$ \\
        4 & 0 &  $-$& $+$ \\
        \hline
        \hline
     
        \end{tabular*}
 
           \begin{tabular*}{0.9\linewidth}{@{\extracolsep{\fill}}c c c c}
        GS & $\langle Z_{v_A} \rangle$ & $\langle Z_{v_B} \rangle$ & $\langle X_{v_A} - X_{v_B} \rangle$  \\
        \hline
        1 & $+$&  $+$ & 0 \\
        2 & $-$&  $+$ & 0 \\
        3 & $+$&  $-$ & 0 \\
        4 & $-$&  $-$ & 0 \\
        \hline
        \end{tabular*}

\end{table}

\subsubsection*{Line(II): charge condensation and SSB via  $X_{e}$-field} \label{subsec:SSBtc}

We now examine the following Hamiltonian:
\begin{equation}
H_{SET} - h^{X}_e\sum_{e}X_e.
\end{equation}
Analogous to the case of regular toric code \cite{trebst2007breakdown,Vidal09A, Tupitsyn10,dusuel2011robustness,Wu12}, this perturbation drives the SET phase into a trivial topological phase by condensing the charge. In Fig.~\ref{fig:L123b}(c) we can see that $\mathcal{O}^{[1,1]}_\ell$ gets sharper with increasing $\ell$ and that the value correctly approaches $-1$ in the whole SET phase. The condensation of the charge is also indicated by the saturation of $\langle X_e \rangle $ in Fig. \ref{fig:L123b}(d).

The condensation of the charge is accompanied by the SSB of the symmetries that fractionalize this excitation. As in Sec.\ref{sec:Hinter}, we find that all symmetries are spontaneously broken, as indicated by the non-zero SSB order parameters.
To better understand the SSB pattern, we can obtain the other SSB ground states by starting with one ground state and applying all possible symmetry actions generated by $X_{A/B}$ and inversion. Doing this, we obtain a total of four orthogonal ground states, with each ground state being invariant under either $X_A$ or $X_B$, but never both. It is interesting to compare this SSB pattern of the ground states with another model with the same symmetries. We consider two 2D Ising models placed on each sublattice of the hexagonal lattice (\textit{i.e.} next-nearest neighbour $ZZ$ Ising interactions on the hexagonal lattice), and we denote this model as ${\rm Ising}_A\otimes {\rm Ising}_B$. This model is invariant under the same symmetries on the vertex spins of $H_{SET}-h^X_e\sum_e X_e$ and moreover, it also breaks all of them spontaneously, resulting in 4 degenerate ground states. However, the four ground states have a different SSB pattern as seen by the SSB order parameters in Table \ref{tab:ssb}. The unusual SSB pattern of the $H_{SET}$ under a $X_e$ field, comparing to the one of ${\rm Ising}_A\otimes {\rm Ising}_B$, could be due to the fact that the SSB is induced by the condensation of an excitation which fractionalizes on-site and inversion symmetries in a joint manner.

By studying the behavior of the local order parameters across the transition, we can see that it is second order, and we can extract an estimate of the corresponding critical exponents, as shown in Fig. \ref{fig:L123b}(d-Inset). We see that $\langle Z_B \rangle$, which indicates the breaking of $X_B$, has a different critical exponent than $\langle X_{v_A} - X_{v_B} \rangle$, the order parameter for inversion symmetry breaking.

% \pepe{The phase transition point is around $h_e^X=0.94$ which is a bigger value than the one corresponding to the regular toric code in a hexagonal lattice [REF?]. This could be interpreted as symmetry protection of the topological order under $X$-field such that coupling the edges give the vertices gives more robustness to the edges... }

\subsubsection*{ Line(III): combination of $X_e-Z_e$ fields
}

The condensation of the charge and flux drives the SET in two different topologically trivial phases. In the former, the trivial phase manifests SSB while the latter does not. This shows a clear difference from the regular toric code in parallel magnetic fields where both condensations end up in the same trivial phase and there is a finite first order line between them \cite{Vidal09A, Wu12}. Here we study the one-parameter Hamiltonian
\begin{equation}
H_{SET} - (2-h^{Z}_e)\sum_{e}X_e  - h^{Z}_e\sum_{e}Z_e,    
\end{equation}
that describes line(III). Along that line, the SOP shows a jump that is indicative of a first order phase transition (Fig. \ref{fig:L123b}(e)). Furthermore, expectation values of on-site magnetizations on the edges and vertices also exhibit a behavior that is consistent with first order phase transition (Fig. \ref{fig:L123b}(f)).

\subsection{ $Z_vZ_{v'}$-interaction: transition to a SSB toric code}

In this section, we analyze the effect of the nearest-neighbour Ising interaction between vertices, {\it i.e.}
\begin{equation} \label{eq:pertising}
H_{SET} - J^{ZZ}_v\sum_{\langle v,v'\rangle }Z_vZ_{v'}.
\end{equation}
The $ZZ$ interaction on the vertices of the honeycomb lattice does not preserve the whole global symmetry, it explicitly breaks part of the symmetry. It only commutes with TRS, inversion, and a $\mathbb{Z}_2$ subgroup of the global $\mathbb{Z}_2\times \mathbb{Z}_2$ (the one generated by acting with $X$ on all vertices). These symmetries still fractionalize the charge, so we expect the symmetry fractionalization to persist as the interaction strength is increased, up to the phase transition.

\begin{figure}[t!]
	\includegraphics[width=0.8\linewidth]{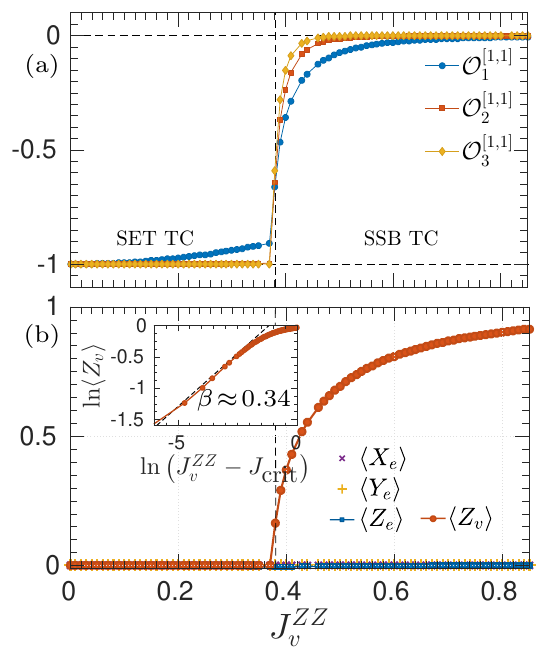}
	\caption{Second order phase transition from the SET phase to a toric code with SSB driven by the Ising field. (a) Values of $\mathcal{O}_\ell^{[1,1]}$, see  Eq.\eqref{eq:opgenerl}, for different block lengths $\ell$. (b)  Magnetization per site along the different field directions on the edges and vertices of the honeycomb lattice. The inset shows the critical exponent $\beta$ of the transition.}
	\label{fig:ising_field}
\end{figure}

The physics of the resulting phase transition are easy to describe, since the $ZZ$ interaction again commutes with $U_{CCZ}$, as in the case of Eq.~\ref{eq:pertz}. If we conjugate Eq.~\ref{eq:pertising} by $U_{CCZ}$, we are left with $H_{TC}$ on the edge spins, and a 2D transverse-field Ising model on the vertices. Therefore, the physics of this transition should be as in this Ising model. Indeed, the local order parameter $\langle Z_v\rangle$ indicates the SSB, and the corresponding critical exponent is in agreement with the 3D classical Ising universality class [see inset in Fig. \ref{fig:ising_field}(b)], although the critical point is slightly lower than what is given in Ref.~\cite{Blote2002}, which we expect to be a result of our relatively low bond dimension.

The large $J_v^{ZZ}$ limit of Eq.~\ref{eq:pertising} will be equivalent to a toric code on the edges with one of the SSB ground states of the Ising model on the vertices, which is consistent with the small values of $\langle X_e \rangle$, $\langle Y_e \rangle$, and $\langle Z_e \rangle$.

Despite the equivalence of Eq.~\ref{eq:pertising} to the Ising model, the SOPs still show non-trivial behaviour, as seen in Fig. \ref{fig:ising_field}(a). We see that, with increasing $\ell$, $\mathcal{{O}}_\ell^{[1,1]}$ approaches -1 in the whole SET phase, indicating that the symmetry fractionalization does indeed remain non-trivial despite partially breaking the on-site symmetry. While the order parameter is not designed to be used when the symmetry is spontaneously broken, it is easy to show that $\mathcal{O}^{[a,b]} = 0$ for all $[a,b]\neq[0,0]$ in the limit $J^{ZZ}_v \rightarrow \infty$, and this is consistent with our results.

\subsection{$Y_{e}$-field}
\begin{figure}[t!]
\includegraphics[width=\linewidth]{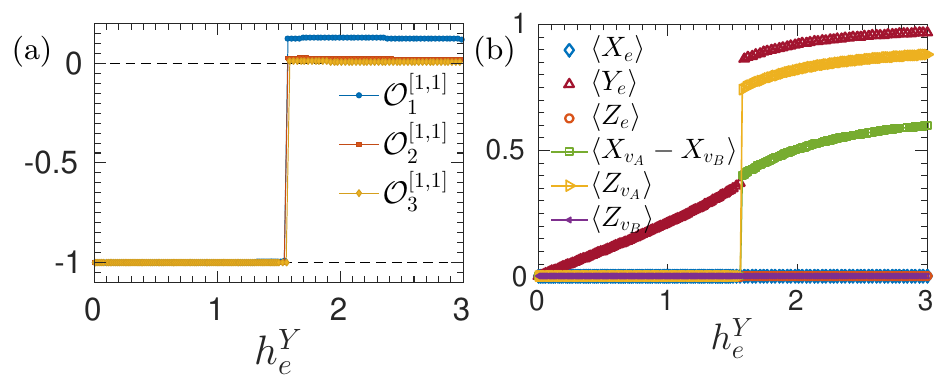}
\caption{ (a) Behaviour of the $\mathcal{O}_l^{[1,1]}$ for different $\ell$ under the action of $Y_e$-field (b) Different magnetizations for the $Y_e$-field perturbation. }
\label{fig:resultsY}
\end{figure}

We now study the effects of $Y$-field,
\begin{equation}
 H_{SET} - h^{Y}_e\sum_{e}Y_e.   
\end{equation}
In \cite{Vidal09B} the authors studied the regular toric code in transverse $Y$-field and found a first order phase transition to a fully polarized phase. Here, with SET order, we also find a first order phase transition to a phase where the edge spins are polarized in the $Y$-direction, as indicated by the discontinuities in the SOP and the local magnetizations, see Fig.\ref{fig:resultsY}. 

Interestingly, this phase transition is also accompanied by SSB, as indicated by the non-zero SSB order parameters. The SSB pattern is the same as observed in Sec.~\ref{subsec:SSBtc}. Unlike the phase transitions induced by parallel $X$ or $Z$ fields, this transition cannot be interpreted as a condensation transition, so we cannot use the general arguments of Section \ref{sec:ssb} to understand the presence of SSB. Nevertheless, it remains possible that the charge excitation is condensed (in the sense of condition \eqref{hy2:charcond} in Section \ref{sec:ssbinT}), so our proof of SSB presented in Appendix \ref{sec:OPandSSB} may still apply. We leave a more detailed analysis of the nature of this transition to future work.

\begin{table*}[]
\caption{\label{tab:table} Summary of the effect and characterization of the different perturbations on $H_{SET}$. All symmetries stand for $\mathbb{Z}_2\times\mathbb{Z}_2 $, inversion (i.e., reflection across bonds), and time-reversal symmetry. All broken stands for the breaking of all the symmetries that were present before the phase transition (this does not include the ones that are explicitly broken by the perturbation).}
\begin{tabular*}{\linewidth}{c @{\extracolsep{\fill}}c @{\extracolsep{\fill}}c @{\extracolsep{\fill}}c}
\hline \hline
Perturbation & Transition(s) & Symmetries of GS(s)  \\
\hline 
$X_e$ & SSB from charge condensation & {All} $\rightarrow$ {{All broken}}  \\ 
$Z_e$ & Flux condensation, no SSB & All $\rightarrow$  All, (No SSB, GS is unique)  \\
$Y_e$  & SSB & {$ (\mathbb{Z}_2\times\mathbb{Z}_2)$, Inversion} $\rightarrow$ {All broken}  \\
$Z_v Z_{v'}$  & Ising SSB & {$\mathbb{Z}_2,$ Inversion, and TRS} $\rightarrow$ {All broken}  \\
 \hline \hline
\end{tabular*}
\end{table*}

\section{Discussion}\label{sec:conclu}

For all of the perturbations considered in the previous section, we see that the string order parameter correctly captures the symmetry fractionalization pattern in the SET phase. Near a phase transition, the use of longer strings is necessary to obtain the fixed-point value of the order parameter. It is remarkable that the order parameters work so well even away from the fixed-point, given that they are defined in terms of the anyon creation operator $S_\Gamma$ which is only exact at the fixed-point. Our results provide evidence that string order gives a ubiquitous characterization of symmetry fractionalization in SET phases.

There are many avenues for generalization of these results. First, Ref.~\cite{Garre19} shows how to construct  string order parameters for detecting other types of symmetry fractionalization in PEPS, and our strategy of lifting these order parameters to physical observables should apply equally well to these examples. Second, it should be possible to use similar methods to study perturbations of any fixed-point SET model where the anyon creation string operators are known  \cite{Mesaros13,Heinrich16,Cheng17}. Going beyond this, we believe that our approach should be applicable even to models for which a fixed-point is not known. As shown by our results, it is not necessary to use the exact anyon operator to detect the symmetry fractionalization in a particular ground state. Therefore, is should be possible to also apply our methods to systems where one can formulate an approximation to the exact anyon operator. For example, one could use an Ansatz like resonating valence bond states, which form good approximations to ground states for certain spin liquids \cite{simplexrvb1,simplexrvb2}, to define approximate anyon operators that can then be used to define string order parameters for these spin liquids.

We emphasize the importance of using SOPs designed for SET phases, as in the present study. This is because there are SF patterns that cannot be characterized just by 1D SPT order parameters \cite{Huang14}. Two examples related to toric code topological order are worth mentioning. First, two of the non-trivial SET phases of a $\mathbb{Z}_2\times \mathbb{Z}_2$ global symmetry cannot be distinguished by 1D SPT order parameters. Second, the non-trivial SF pattern given by a global $\mathbb{Z}_2$ symmetry has no analogous 1D SPT phase; this is the case of the model studied in \cite{Hermele14} and the ones in \cref{app:TCPEPS}. Our methods are able to characterize all of these phases \cite{Garre19}. We hope our work motivates the use of these SOPs to study unknown phases of matter in 2D.

Apart from the string order parameters, we have also observed several instances of SSB across the various phase transitions, as summarized in Table \ref{tab:table}. The SSB induced by the $X_e$ field can be explained by the condensation of an anyon that fractionalizes the symmetry. On the other hand, we also observe SSB in the intermediate phase of the direct Hamiltonian interpolation, and under the transverse $Y_e$ field, despite the fact that neither transition has a clear interpretation has a condensation transition. Furthermore, in all three cases, the symmetry breaking pattern, which involves both the on-site and inversion symmetries, is distinct from that of a pair of Ising models which share the same symmetries. These findings suggest that there is yet more to uncover about the connection between topological phase transitions and spontaneous symmetry breaking.

\section*{Acknowledgements}
J.G.R would like to acknowledge Sofyan Iblisdir for interesting discussions and for his invaluable support. MI and DTS would like to thank Norbert Schuch for helpful comments. This work has received funding from the European Research Council (ERC) under the European Union’s Horizon 2020 research and innovation program through Grant Nos. 636201 WASCOSYS, and 648913 GAPS. DTS was supported by a fellowship from the Natural Sciences and Engineering Research Council of Canada (NSERC), and by the Deutsche Forschungsgemeinschaft (DFG) under Germany’s Excellence Strategy (EXC-2111 – 390814868). Numerical computations have been performed on the TQO cluster of Max-Planck-Institute of Quantum Optics. 

\appendix

\section{PEPS for the GS of $H_{SET}$}\label{sec:TNGS}

The ground state of $H_{SET}$ has a tensor network description. We divide the lattice into the two types of vertices, $A$ and $B$. The PEPS is constructed by assigning different tensors to each type of vertex and to the edges. The resulting PEPS has bond dimension 3. The non-zero components of the tensor of the edges, $T_e$, are: 
\begin{equation}
1=
\begin{tikzpicture}[scale=0.5]
    \node [anchor = east,scale=2] at (-1,0) {{$2$}};
 \node [anchor = west, scale=2] at (1,0) {$2$};
 \draw (-1,0)--(1,0);
    \draw[thick, fill=white] (0.0,0.0) circle (0.1cm);
 \node [anchor = south,scale=2] at (0,0.10) {$0$};
\end{tikzpicture},\
1 =
\begin{tikzpicture}[scale=0.5]
    \node [anchor = east,scale=2] at (-1,0) {$0$};
 \node [anchor = west,scale=2] at (1,0) {$0$};
\draw (-1,0)--(1,0);
\draw[thick, fill=white]  (0,0) circle (0.1cm);
 \node [anchor = south,scale=2] at (0,0.10) {$1$};
\end{tikzpicture}
{=}
\begin{tikzpicture}[scale=0.5]
    \node [anchor = east,scale=2] at (-1,0) {$1$};
 \node [anchor = west,scale=2] at (1,0) {$1$};
\draw (-1,0)--(1,0);
\draw[thick, fill=white] (0,0) circle (0.1cm);
 \node [anchor = south,scale=2] at (0,0.10) {$1$};
\end{tikzpicture}.
\end{equation}
This tensor projects the virtual legs on each edge onto the state labelled 2 if there is no loop running along that edge, and otherwise onto the span of the states labelled by 0 and 1. 

The non-zero components of the tensors of vertices $A$ and $B$, $T_A$ and $T_B$, are respectively 
\begin{equation}
1=
\begin{tikzpicture}[scale=0.5]
 \node [anchor = west, scale=2] at (1.45,0) {$2$};
  \node [scale=2] at (120:1.6) {$2$};
\draw[scale=.9] (1.45,0)--(0,0)-- (120:1.48);
\draw[thick] (0,0)-- (60:0.3);
 \node [scale=2] at (55:0.6) {$0,1$};
  \node[scale=2] at (0.5,-0.5) {  $T_A$ };
\draw[scale=.9] (0,0)-- (-120:1.5);
  \node [scale=2] at (-120:1.65) {$2$};
\end{tikzpicture}
,\ 1=
\begin{tikzpicture}[scale=0.5]
  \node[scale=2] at (0.5,-0.5) {  $T_A$ };
 \node [anchor = west,scale=2] at (1.5,0) {$-$};
  \node [scale=2] at (120:1.6) {$-$};
\draw[scale=.9] (1.5,0)--(0,0)-- (120:1.5);
\draw[thick,scale=.9] (0,0)-- (60:0.3);
 \node [scale=2] at (55:0.6) {$1$};
\draw[scale=.9] (0,0)-- (-120:1.5);
  \node [scale=2] at (-120:1.65) {$2$};
\end{tikzpicture}
=
\begin{tikzpicture}[scale=0.5]
  \node[scale=2] at (0.5,-0.5) {  $T_A$ };
 \node [anchor = west,scale=2] at (1.5,0) {$+$};
  \node [scale=2] at (120:1.6) {$+$};
\draw [scale=.9] (1.5,0)--(0,0)-- (120:1.5);
\draw[thick,scale=.9] (0,0)-- (60:0.3);
 \node [scale=2] at (55:0.6) {$0$};
\draw[scale=.9] (0,0)-- (-120:1.5);
  \node [scale=2] at (-120:1.65) {$2$};
 \end{tikzpicture},
\end{equation}

 \begin{equation}
 1=
 \begin{tikzpicture}[scale=0.5]
   \node[scale=2] at (-0.5,-0.5) {  $T_B$ };
 \node [anchor = east,scale=2] at (-1.5,0) {$2$};
   \node [scale=2] at (60:1.6) {$2$};
 \node [scale=2] at (120:0.45) {$0,1$};
\draw [scale=0.9] (-1.5,0)--(0,0)-- (60:1.5);
\draw [thick] (0,0)-- (120:0.3);
\draw [scale=0.9] (0,0)-- (-60:1.5);
 \node [scale=2] at (-60:1.6) {$2$};
\end{tikzpicture}
=
\begin{tikzpicture}[scale=0.5]
   \node[scale=2] at (-0.5,-0.5) {  $T_B$ };
 \node [anchor = east,scale=2] at (-1.5,0) {$0$};
   \node [scale=2] at (60:1.6) {$0$};
 \node [scale=2] at (120:0.45) {$0$};
\draw [scale=0.9] (-1.5,0)--(0,0)-- (60:1.5);
\draw[thick, scale=0.9] (0,0)-- (120:0.3);
\draw [scale=0.9] (0,0)-- (-60:1.5);
 \node [scale=2] at (-60:1.6) {$2$};
\end{tikzpicture}
=
\begin{tikzpicture}[scale=0.5]
   \node[scale=2] at (-0.5,-0.5) {  $T_B$ };
 \node [anchor = east,scale=2] at (-1.5,0) {$1$};
    \node [scale=2] at (60:1.6) {$1$};
 \node [scale=2] at (120:0.45) {$1$};
\draw [scale=0.9] (-1.5,0)--(0,0)-- (60:1.5);
\draw[scale=0.9,thick] (0,0)-- (120:0.3);
 \node [scale=2] at (-60:1.6) {$2$};
\draw [scale=0.9] (0,0)-- (-60:1.5);
\end{tikzpicture},
\end{equation}
plus all their rotations where $\pm=|0\rangle \pm |1\rangle$. The first role of the vertex tensors is to enforce an even number of virtual $|2\rangle$ states at each vertex which gives the closed-loop condition. Then, the rest of the structure serves to create the decoration by cluster states. This can be seen by imagining removing the legs labelled 2 in the above equations. Then, the resulting tensors resemble those defining the 1D cluster state \cite{Gross07}. 

The symmetries of these tensors are the following:
\begin{align}
X T_A =& \; T_A (Z\oplus 1)^{\otimes 3}, & \; T_A(X\oplus 1)^{\otimes 3}= T_A, \nonumber \\
X T_B = & \; T_B (X\oplus 1)^{\otimes 3}, &\; T_B(Z\oplus 1)^{\otimes 3}= T_B, \nonumber \\
T_A(\bar{Z})^{\otimes 3}& = T_A,& \; T_B(\bar{Z})^{\otimes 3}= T_B,
\end{align}
where $\bar{Z}= -1\oplus -1\oplus 1$. 

We block the hexagonal lattice to a square one by contracting $T_A$, $T_e$ and $T_B$ (independent of the direction) to result in the tensor $T$: 
\begin{equation} \label{eq:blockedtensor}
T=
\begin{tikzpicture}[scale=0.5]
\draw[] (-120:1.5)--(0,0)-- (120:1.5);
\draw[thick] (0,0)-- (60:0.3);
\draw[dotted] (1,1.5)-- (1,-1.5);
  \node[scale=1.5] at (0,0.5) { $v_A$}; \node[scale=1.5] at (2,0.5) { $v_B$};
  \node[scale=1.5] at (1,-0.5) { $e_2$};
  \node[scale=1.5] at (-90:0.75) {$e_1$};
    \node[scale=1.5] at ($(2,0)+(-90:0.75)$) {$e_3$};
  \draw[] (0,0)-- (2,0);
  \draw[thick] (2,0)--($(2,0)+(120:0.3)$);
  \draw[] ($(2,0)+(60:1.5)$)--(2,0)-- ($(2,0)+(-60:1.5)$);
   \draw[thick, fill=white] (1,0.0) circle (0.1cm);
    \draw[thick, fill=white] (-120:0.75) circle (0.1cm);
    \draw[thick, fill=white] ($(2,0)+(-60:0.75)$) circle (0.1cm);
  \end{tikzpicture}  .
\end{equation}
This tensor has the following symmetries:
\begin{align}
T = & \; T\bar{Z}^{\otimes 4}, \nonumber \\
X_{v_B} T =  & \;T (X\oplus 1)^{\otimes 4}, \nonumber \\
X_{v_A} T = &\;  T (Z\oplus 1)^{\otimes 4}, \nonumber \\
X_{v_A}\otimes X_{v_B} T = &\; T (iY\oplus 1)^{\otimes 2} \otimes  (-iY\oplus 1)^{\otimes 2},
\end{align}
where the first equation accounts for the $\mathbb{Z}_2$-injectivity. Then, the state has a global symmetry corresponding to the group $\mathbb{Z}_2\times \mathbb{Z}_2$. The SF pattern is given in \eqref{SFenrtc} which corresponds to the $D_8$ gauge theory (see \cite{Garre19}). If we denote by $v_q$, $q\in \mathbb{Z}_2 \times \mathbb{Z}_2 $ the virtual symmetry operators that satisfy $v_q v_k = u(q,k) v_{qk}$ we find that the matrices $u(q,k)$ are
\begin{equation}\label{SFpatterntc} u_{\omega(q,k)}=
  \left( {\begin{array}{cccc}
 {\color{blue}\id} & \id &  {\color{blue}\id} &\id \\
 \id & \id & \bar{Z} & \bar{Z} \\
 {\color{blue}\id} & \id &  {\color{blue}\bar{Z}} & \bar{Z} \\
 \id & \id & \id&\id 
  \end{array} } \right)_{q,k},
\end{equation}
where they are ordered as $X_{v_B},X_{v_A}\otimes X_{v_B},X_{v_A}$. The blue cells colored in Eq.\eqref{SFpatterntc} correspond to the unique $\mathbb{Z}_2$  subgroup (on-site symmetry $X_{v_A}\otimes X_{v_B}$) of the global symmetry whose restriction results in a non-trivial SF pattern.

Let us show how bond-centered inversion on the hexagonal lattice is represented in $T$. The action is given by a permutation of the virtual legs in $T$ across the inversion axis (dotted line in Eq.~\ref{eq:blockedtensor}) together with a physical swap of the vertices $v_{A/B}$ and edges $e_{1/3}$. This action turns out to be equal to a Hadamard gate $H=\frac{X+Z}{\sqrt{2}}$ acting on each virtual leg so inversion transforms the tensor as $T \rightarrow T H^{\otimes 4}$. This transformation allows us to compute the interplay between the virtual operators of the global $\mathbb{Z}_2 \times \mathbb{Z}_2 $ symmetry and inversion symmetry. For example, the anticommutation relation of $iY$ and $H$ reveals this nontrivial interplay.

%I XAXB I XA I XB I is non-trivial since HiYH x HXH = -1

\section{SET on PEPS, order parameters and SSB} \label{sec:OPandSSB}
%%%
In this section, we first introduce PEPS and show how topological order, global symmetries, and their interplay are characterized. Then, we elaborate on the order parameter that we have used in the main text to study the SET phases. Finally, we prove the emergence of SSB when an  anyon that transforms non-trivially under the symmetry is condensing.

\subsection{Background of topological order and global symmetries on PEPS}

\begin{figure}[ht!]
\begin{center}
\includegraphics[scale=1.2]{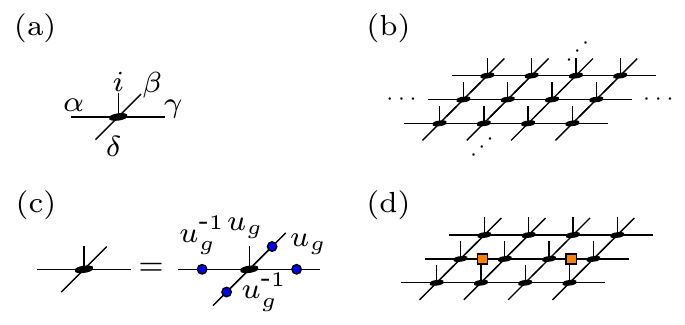}
\caption{(a) Diagram for the PEPS tensor $A^i_{\alpha \beta \gamma \delta}$ on the square lattice. (b) The PEPS constructed via the contraction of the virtual d.o.f. of the tensors. The contraction of indices is represented by joining the corresponding legs associated to the indices. We place the PEPS on a torus however, we will leave the boundaries open in the drawings for the sake of clarity. (c) Invariance of the $G$-injective tensor under the action of the group $G$  on the virtual d.o.f. (d) Pair of charge operators placed on the virtual d.o.f.}
\label{fig:PEPSconstruction}
\end{center}
\end{figure}

PEPS are pure states completely characterized by a tensor $A$. We focus on bosonic and translational invariant systems on the square lattice for the sake of simplicity. The tensor $A^i_{\alpha,\beta,\gamma,\delta}$--see Fig.\ref{fig:PEPSconstruction}(a)--then has five indices; one $i=1,\dots,d$ for  the physical Hilbert space $\mathbb{C}^d$ of each particle and four $\alpha,\beta,\gamma,\delta= 1,\dots, D$ which correspond to the virtual degrees of freedom (d.o.f.). The PEPS is constructed by placing $A$ on each vertex and contracting the neighbour virtual d.o.f. (identifying and summing the indices) as depicted in Fig.\ref{fig:PEPSconstruction}(b). When the chosen boundary conditions are applied, the resulting tensor contraction $c_{i_1,\cdots,i_N}=\mathcal{C}\{ A^{i_1},\dots, A^{i_N}\}$ describes a quantum many-body state $|\psi_A\rangle =\sum c_{i_1,\cdots, i_N}|i_1\cdots i_N\rangle$. 

\subsubsection{${G}$-injective PEPS and global symmetries}
%%%
We focus on the family of $G$-injective PEPS \cite{Schuch10} whose tensors have the following virtual symmetry ($G$-invariance) illustrated in Fig.\ref{fig:PEPSconstruction}(c):
\begin{equation}
A=A(u_g \otimes u_g \otimes \myinv{u}_g \otimes \myinv{u}_g ),
\end{equation}
where here $A$ is represented as a map from the virtual to physical indices, and $u_g$ is some unitary representation of the finite group $G$. Given a $G$-injective PEPS, the associated parent Hamiltonian \cite{Schuch10}, defined on a torus, has the ground state degeneracy $\mathcal{D}$($G$) of  the quantum double model of $G$.  In this work, we focus on abelian $G$ for the sake of simplicity.

The anyons of $\mathcal{D}$($G$) are charges, fluxes, and dyons (a combined excitation of the previous two). Charges are labelled by the irreducible representations (irreps) of $G$ and fluxes by elements of $G$. We represent the virtual operator of a pair of charges, $\Pi_{\sigma}=C_{\sigma}\otimes C_{\bar{\sigma}}$, as two orange rectangles placed on two different edges of the lattice, see Fig.\ref{fig:PEPSconstruction}(d).

We consider $G$-injective PEPS,  $|\psi_A\rangle$, with a global on-site symmetry 
\begin{equation}
U_q^{\otimes n}|\psi_A\rangle=|\psi_A\rangle \;\forall q\in Q,
\end{equation}
where $U_q$ is a linear unitary representation of some finite group $Q$ and $n$ is the number of lattice sites. 
For all $ q \in Q $, there is an invertible matrix $v_q$ which translates $U_q$ through the local tensor $A$ on the virtual d.o.f. as follows \cite{Garrethesis}: 
 
  \begin{equation} \label{locsym}
    \begin{tikzpicture}[baseline=-1mm, scale=1.2]
      \pic[thick] at (0,0,0) {3dpeps};
      \draw[thick] (0,0,0) -- (0,0.35,0);
      \filldraw[draw=black,fill=red,thick] (0,0.2,0) circle (0.07);
      \node[anchor=east] at (0,0.3,0) {$U_q$};
    \end{tikzpicture} =
    \begin{tikzpicture}[scale=1.2]
      \draw[thick] (-0.7,0,0) -- (0.7,0,0);
      \draw[thick] (0,0,-0.9) -- (0,0,0.9);
      \pic[thick] at (0,0,0) {3dpeps};
      \filldraw[draw=black,fill=purple,thick] (-0.5,0,0) circle (0.06);
      \filldraw[draw=black,fill=purple,thick] (0.5,0,0) circle (0.06);
      \node[anchor=south] at (-0.5,0,0) {$\myinv{v_q}$};
      \node[anchor=north] at (0.5,0,0) {$v_q$};
      \filldraw[draw=black,fill=purple,thick] (0,0,-0.6) circle (0.06);
      \filldraw[draw=black,fill=purple,thick] (0,0,0.6) circle (0.06);
       \node[anchor=south] at (0,0,-0.6) {$v_q$};
      \node[anchor=north] at (0,0,0.6) {$\myinv{v_q}$};
    \end{tikzpicture}
    \; \forall q\in Q.
  \end{equation}
 The operators $v_q$ do not have to form a linear representation. It actually turns out that 
\begin{equation}
    v_k v_q = u_{\omega(k,q)} \; v_{kq},
\end{equation}
where $\omega(k,q) \in G$. This means that $\{ v_q \}$ form a homomorphism up to the matrix $u_{\omega(k,q)}$, where $u$ is the representation of $G$ introduced earlier for the $G$-invariance. 

The action of the global symmetry on a charge sitting on a virtual bond is given by $C_{\sigma} \to \Phi_q(C_{\sigma})$, where $\Phi_q( X ) = v_q X v^{-1}_q$. Diagrammatically: 
\begin{equation}
  \begin{tikzpicture}[scale=1.2]
   \pic at (0,0,0.7) {3dpepsshort}; 
        \pic at (0,0,2.1) {3dpepsshort};
        \pic at (0,0,1.4) {3dpepsshort}; 
      \pic at (0.5,0,0.7) {3dpepsshort};
      \pic at (0.5,0,1.4) {3dpepsshort};
     \pic at (0.5,0,2.1) {3dpepsshort};
     	 \filldraw[draw=black,fill=red,thick] (0,0.11,0.7) circle (0.05);
	 \filldraw[draw=black,fill=red,thick] (0,0.11,2.1) circle (0.05);
	 \filldraw[draw=black,fill=red,thick] (0,0.11,1.4) circle (0.05);
	 \filldraw[draw=black,fill=red,thick] (0.5,0.11,0.7) circle (0.05);
	 \filldraw[draw=black,fill=red,thick] (0.5,0.11,1.4) circle (0.05);
	 \filldraw[draw=black,fill=red,thick] (0.5,0.11,2.1) circle (0.05);
	    \node[anchor=south] at (-0.1,0.15,1)  {$U_q$};
	         \filldraw[draw=black,fill=orange,thick] (0.15,0,1.27) rectangle (0.35,0,1.53); 
 \end{tikzpicture} 
 =
 \begin{tikzpicture}[scale=1.2]
     \pic at (0,0,0.7) {3dpepsshort}; 
        \pic at (0,0,2.1) {3dpepsshort};
        \pic at (0,0,1.4) {3dpepsshort}; 
      \pic at (0.5,0,0.7) {3dpepsshort};
      \pic at (0.5,0,1.4) {3dpepsshort};
     \pic at (0.5,0,2.1) {3dpepsshort};

         \filldraw[draw=black,fill=purple,thick]  (0.12,0,1.4) circle (0.04);
	  \filldraw[draw=black,fill=orange,thick] (0.15,0,1.27) rectangle (0.35,0,1.53); 
	\filldraw[draw=black,fill=purple,thick]  (0.38,0,1.4) circle (0.04);
 \end{tikzpicture} 
,
  \begin{tikzpicture}[scale=1.2]
      \draw[thick] (0.1,0,0) -- (0.7,0,0);
      \filldraw[draw=black,fill=purple,thick] (0.2,0,0) circle (0.05);
      \filldraw[draw=black,fill=purple,thick] (0.6,0,0) circle (0.05);
      \node[anchor=south] at (0.2,0,0) {$v_q$};
      \node[anchor=north] at (0.65,0,-0.15) {$\myinv{v_q}$};
       \filldraw[draw=black,fill=orange,thick] (0.4,0,-0.2) rectangle (0.4,0,0.2); 
    \end{tikzpicture}
        \equiv  \Phi_q(C_{\sigma}).
 \end{equation}
If the symmetry is applied for two elements $q,k\in Q$, we see that  
\begin{equation}\label{eq:symcharge}
(\Phi_k\circ \Phi_q)(C_{\sigma})=(\tau_{\omega(k,q)}\circ \Phi_{kq})(C_{\sigma}),
\end{equation}
where $\tau_{\omega(k,q)}$ denotes the conjugation by $u_{\omega(k,q)}$. This implies that the symmetry action over the charge sector can be {\it projective}, {\it i.e.} the symmetry fractionalizes. 
Let us assume that $u_g v_q = v_qu_g$ for all $g\in G$ and $q\in Q$ which means that the global symmetry does not permute between anyons. Formally we define $\omega : Q \times Q \to G: (k,q) \to \omega(k,q) = v_k v_q v^{-1}_{kq}$ that satisfies a 2-cocycle condition. The possible 2-cocycles are classified by the second cohomology group $H^2(Q,G)$ which characterizes the different SF patterns of $G$-injective PEPS. 
\begin{comment}

\subsubsection{Inversion symmetry}
We consider now the case of a $G$-injective PEPS invariant under reflection with respect to a horizontal line. This symmetry is realized at the level of tensors by
$$ U_\sigma A \pi=A(\myinv{V}\otimes W\otimes V\otimes \myinv{W}),$$
where $U_\sigma$ is a transposition of the blocked sites of the tensors, $\pi$ is the horizontal flip operator in the virtual level (interchange plus transposition) and $V$ is the virtual operator acting on the horizontal part. Notice that we are assuming a translational invariance under blocked tensors. If we apply again another horizontal reflection, it follows that
\begin{align*}
A= & \;U_\sigma A(\myinv{V}\otimes W\otimes V\otimes \myinv{W}) \pi\\
 =& \; [U_\sigma A\pi](\myinv{V}\otimes (\myinv{W})^T \otimes V\otimes W^T )\\
 =&  \; A(V^{-2}\otimes W(\myinv{W})^T \otimes V^2\otimes \myinv{W} W^T ),
 \end{align*}
which implies that $V^2=W(\myinv{W})^T\in G$ .
\end{comment}
\subsection{Order parameters for symmetry fractionalization}
%%%

In Ref.\cite{Garre19} the authors constructed order parameters that identify the SF pattern of the charges without relying on the knowledge of the virtual symmetry operators $v_q$ for the RGFP of $G$-injective PEPS. Here, we  extend their definition to $G$-injective PEPS where a perturbation is applied.

We consider on-site perturbations of the tensor $A$ of the form $A(\theta)=T(\theta)A$, where $T(\theta)$ is some invertible matrix that depends on $\theta$. We notice that this kind of perturbation keeps the $G$-injectivity of the tensor. Then, the virtual charge operator can be the same as in the RGFP.

For $G=\mathbb{Z}_2$, $Q=\mathbb{Z}_2=\{e,a\}$, the order parameter that we propose to capture the SF pattern is the following:

\begin{equation}\label{transprotocol}
\mathcal{O}^{[a]}= \frac{
 %\langle \psi_{A(\theta)}|\Lambda^{[a]}|\psi_{A(\theta)}\rangle 
 \mathcal{L}^{[a]}(\theta)}{
 %\langle \psi_{A(\theta)}|\Lambda^{[e]}|\psi_{A(\theta)}\rangle 
 \mathcal{L}^{[e]}(\theta)}.
\end{equation}
The value of  $\mathcal{L}^{[a]}(\theta)$ is given by 
\begin{equation}
\begin{tikzpicture}[scale=1]
     \pic at (0,0,0.7) {3dpepsshort};
    \pic at (0,1.5,0.7) {3dpepsdownshort};
    \pic at (0.5,0,0) {3dpeps};
    \pic at (0.5,1.5,0) {3dpepsdown};
    \pic at (0.5,0,1.4) {3dpeps};
    \pic at (0.5,1.5,1.4) {3dpepsdown};
    \pic at (1,0,0) {3dpeps};
    \pic at (1,1.5,0) {3dpepsdown};
    \pic at (1,0,1.4) {3dpeps};
    \pic at (1.5,0,0.7) {3dpeps};
    \pic at (1.5,1.5,0.7) {3dpepsdown};
	\pic at (2,0,0.7) {3dpepsshort};
    \pic at (2,1.5,0.7) {3dpepsdownshort};
    \begin{scope}[canvas is xy plane at z=0.7]
        \draw[preaction={draw, line width=1.2pt, white},blue,thick] (0.5,1.5,0.7) to  (0.5,1.05,0.7);
        \draw[preaction={draw, line width=1.2pt, white},blue,thick] (1,1.5,0.7) to  (1,1.05,0.7);
        \draw[preaction={draw, line width=1.2pt, white},blue,thick] (1,1.1,0.7) to [out=-90, in=90] (0.5,0,0.7);
        \draw[preaction={draw, line width=1.2pt, white},blue,thick] (0.5,1.1,0.7) to [out=-90, in=90] (1,0,0.7);
    \end{scope}   
    \draw[] (1,1.5,1.4)--(1,1.31,1.4);
    \begin{scope}[canvas is zx plane at y=1.5]
        \draw[preaction={draw, line width=1.2pt, white}] (0.9,1)--(1.9,1);
        \draw[preaction={draw, line width=1.2pt, white}] (1.4,0.6)--(1.4,1.4);
        \filldraw (1.4,1) circle (0.07);
    \end{scope} 
    \pic at (1,0,0.7) {3dpepsp};
    \pic at (1,1.5,0.7) {3dpepsdownp};
    \pic at (0.5,0,0.7) {3dpepsp};
    \pic at (0.5,1.5,0.7) {3dpepsdownp};
    \filldraw[draw=black,fill=orange,thick] (1.25,0,0.5) rectangle (1.25,0,0.9 );
    \filldraw[draw=black,fill=orange,thick] (0.25,1.5,0.5) rectangle (0.25,1.5,0.9); 
    \filldraw[draw=black,fill=orange,thick] (1.75,1.5,0.5) rectangle (1.75,1.5,0.9 ); \filldraw[draw=black,fill=orange,thick] (1.75,0,0.5) rectangle (1.75,0,0.9  );
    \filldraw[draw=black,fill=red,thick] (0.52,0.95,0.7) circle (0.07);
    \node[anchor = east] at (0.52,0.95,0.7) {$U_a$};
    \filldraw[draw=black,fill=red,thick] (0.98,0.95,0.7) circle (0.07);
    \node[anchor = west] at (0.98,0.95,0.7) {$U_a$};
\end{tikzpicture},
\end{equation}
where the depicted tensor is $A(\theta)$ and the blue lines correspond to the permuted sites (also where the symmetry operators act before doing the scalar product). It is important to note that $\mathcal{L}^{[a]}(\theta)$ is not given in terms of an expectation value of an operator by $|\psi_{A(\theta)}\rangle$. This is because of the presence of virtual charge operators.

For $Q=\mathbb{Z}_2\times \mathbb{Z}_2=\{e,a,b,ab\}$ the order parameter is the triple $\{\mathcal{O}^{[a]},\mathcal{O}^{[b]},\mathcal{O}^{[ab]}\}$. The key point is that this order parameter behaves as the fractionalization class of the charge: $\mathcal{O}^{[q]} \approx \omega(q,q)$, that is, it reveals the sign of this action.

The order parameter of Eq.\eqref{transprotocol} is meant for zero correlation length states. However, a perturbation will generally increase the correlation length until the phase transition point is achieved. To account for the growth of the correlation length the following order parameters are defined by blocking sites:
\begin{equation} \label{SOPtnl}
\mathcal{{O}}^{[a]}_{\ell} = \frac{ 
%\langle \psi_{A(\theta)}|\Lambda_{\ell}^{[a]}|\psi_{A(\theta)} \rangle 
\mathcal{L}^{[a]}_\ell(\theta)
}{  
%\langle \psi_{A(\theta)}| \Lambda_{\ell}^{[e]}|\psi_{A(\theta)}\rangle 
\mathcal{L}^{[e]}_\ell(\theta)
}
 \end{equation}
where $\mathcal{L}^{[a]}_\ell(\theta) $ is equal to 
\begin{equation} 
\begin{tikzpicture}[baseline=+1mm][scale=1]
    \pic at (0,0,0.7) {3dpepspb};
    \draw[](0,0,0.7)--(0,1,0.7);
    \pic at (0,1,0.7) {3dpepspb};
    \pic at (4.4,0,0.7) {3dpepspb};
    \draw[](4.4,0,0.7)--(4.4,1,0.7);
    \pic at (4.4,1,0.7) {3dpepspb};  
    \draw[](0,1,0.7)--(4.4,1,0.7);
    \draw[](0,0,0.7)--(4.4,0,0.7);
    \foreach \x in {0.5,0.7,0.9,1.1,1.3}{
        \draw[preaction={draw, line width=1.2pt, white},thick, blue] (\x,1,0.7) to [out=-90, in=90] (1.3+\x,0,0.7);
    }  
    \foreach \x in {0.5,0.7,0.9,1.1,1.3}{
        \draw[preaction={draw, line width=1.2pt, white},thick,blue] (\x+1.3,1,0.7) to [out=-90, in=90] (\x,0,0.7);
        \filldraw[draw=black,fill=red,thick] (\x+0.05,0.8,0.7) circle (0.06);
        \filldraw[draw=black,fill=red,thick] (\x+1.3-0.05,0.8,0.7) circle (0.06);
    }  
    \foreach \x in {0.5,0.7,0.9,1.1,1.3}{
        \pic at (\x,0,0.7) {3dpepsp};
        \pic at (\x,1,0.7) {3dpepsp};
        \pic at (1.3+\x,0,0.7){3dpepsp};
        \pic at (1.3+\x,1,0.7) {3dpepsp};
        \draw[](2.6+\x,0,0.7)--(2.6+\x,1,0.7);
	    \pic at (2.6+\x,0,0.7) {3dpepspb};
        \pic at (2.6+\x,1,0.7) {3dpepsdownpb};
    } 
    \node[anchor = west] at (1.3+1.3-0.05,0.8,0.7) {$U_a$};
    \draw [decorate,decoration={brace,raise=5pt},yshift=0pt]
    (0.4,1,0.7) -- (1.4,1,0.7) node [black,midway,yshift=0.4cm] {\footnotesize $\ell$};
    \filldraw[draw=black,fill=orange,thick] (2.85,0,0.5) rectangle (2.85,0,0.9 );
    \filldraw[draw=black,fill=orange,thick] (0.25,1,0.5) rectangle (0.25,1,0.9); 
  	\filldraw[draw=black,fill=orange,thick] (4.15,0,0.5) rectangle (4.15,0,0.9 );
  	\filldraw[draw=black,fill=orange,thick] (4.15,1,0.5) rectangle (4.15,1,0.9 );
\end{tikzpicture}
\end{equation}
and $\ell$ should be taken greater than the correlation length.

\subsection{SSB from anyon condensation in PEPS} \label{sec:ssbinT}

In this section, we use the framework of PEPS to prove that there is SSB in the ground subspace if a condensed anyon transforms non-trivially under the symmetry, \textit{i.e.} it either fractionalizes or it is permuted. We remark that this has been proven in \cite{Bischoff19} using the language of $G$-graded tensor categories. 

The proof is done by analyzing the fixed point structure of the transfer operator of the PEPS. We show that there is a contradiction if we suppose that the three following conditions hold at the same time: (i) there is no symmetry breaking on the fixed point structure, {\it i.e.} expectation values are independent of the fixed point with whom they are evaluated, (ii) the global symmetry fractionalizes on an anyon $b$ or it permutes $b$ to $c$, (iii) the anyon $b$ condenses or the anyon $b$ condenses and not $c$. To do so we use the framework developed in \cite{Duivenvoorden17,Iqbal18} to describe condensation of anyons in $G$-injective PEPS together with the characterization of symmetries in \cite{Garrethesis} that we revisit now.

\begin{figure}[ht!]
\includegraphics[scale=0.9]{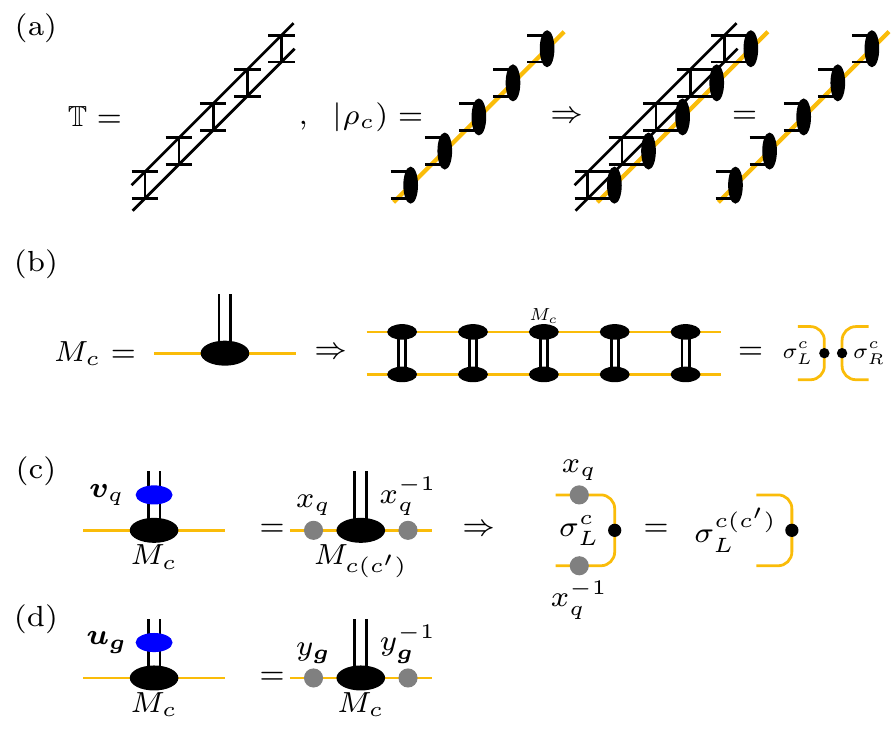}
\caption{ (a) Transfer operator $\mathbb{T}$, it fixed points vectors $|\rho_c )$ and the equation that relates $\mathbb{T}$ and $|\rho_c )$. (b) Any fixed point $|\rho_c )$ is an injective MPS constructed with the tensor $M_c$. This implies the uniqueness of the left/right fixed points, $\sigma^c_{L/R}$, of the transfer matrix constructed with $M_c$. (c) Transformation rule of $M_c$ and $\sigma^c_L$ under $\bm{v}_q$. (d) Transformation rule of $M_c$ under a operator $\bm{u_g}$ corresponding to an unconfined flux, see \cite{Duivenvoorden17}. }
\label{fig:proofobjts}
\end{figure}

In $G$-injective PEPS the transfer operator $\mathbb{T}$ has a degenerate fixed point structure: $ \{|\rho_c)\}$ such that  $\mathbb{T} |\rho_c) = |\rho_c)$, see \cref{fig:proofobjts}(a). This comes from the symmetries of the transfer operator: $[\bm{u_g}^{\otimes L} , \mathbb{T}] = 0$, where $\bm{u_g} \equiv u_g\otimes u_{g'}$ defined for all $\bm{g}  \in G\times G$.  As in Ref.\cite{Duivenvoorden17} we assume that $G$ is abelian, that the fixed point subspace is spanned by injective MPS with tensors $ \{M_c\}$, and that for each $c,c'$ there is a $\bm{g}$ such that  $\bm{u_g}^{\otimes L} |\rho_c) = |\rho_{c'})$ 
%group action of $\{ \bm{u_g} \}$ on  $ \{|\rho_c)\}$ is transitive

A global symmetry on the PEPS is reflected in the transfer operator as follows $ [\bm{v}^{\otimes L}_q,  \mathbb{T}] = 0$, where $\bm{v}_q \equiv v_q\otimes v^{-1}_q$.  The global symmetry acts on the fixed point subspace as follows: 
\begin{equation}
\bm{v}^{\otimes L}_q |\rho_c) = \lambda_q  |\rho_{c'}), \; \forall q\in Q, \tag{0}\label{hy:sympermfp}
\end{equation}
where $|\lambda_q|=1$, $\lambda_q$ does not depend on $c$, and we assume that the symmetry operators can only permute between the fixed points. Since we assumed that $|\rho_c)$ are injective MPS,  the transfer operator constructed with any of these states has a unique right/left fixed point, $\sigma_{R/L}^c$ depicted in \cref{fig:proofobjts}(b). Eq.\eqref{hy:sympermfp}, is translated into how the virtual operators transform those fixed points, see \cref{fig:proofobjts}(c).

The main assumption here is that the expectation values do not depend on the fixed point where they are evaluated:
\begin{equation}
(\rho_c|\mathbb{T}_{\mathcal{O}}|\rho_c) = (\rho_{c'}|\mathbb{T}_{\mathcal{O}}|\rho_{c'}), \; \; \forall c,c' \; \& \;  \mathcal{O},
\tag{I} \label{hy1:nonSB}
\end{equation}
where $\mathcal{O}$ is an observable and $\mathbb{T}_{\mathcal{O}}$ is it virtual representation using the transfer operator of the tensor.
This is a non-symmetry breaking condition since the fixed points are related to the different ground states so that Eq.\eqref{hy1:nonSB} is equivalent to $\langle \mathcal{O} \rangle_c = \langle U_q \mathcal{O} U^\dagger_q \rangle_c$. For example, a SB pattern like the magnetization reads $\langle m_z \rangle_c = -\langle m_z \rangle_{c'}$. In particular, this holds for the evaluation on the fixed points of an anyon operator $[g,\alpha]$ defined as:
\begin{equation}
\begin{tikzpicture}[scale=0.75]
\draw[thick, rounded corners, yellow!50!orange] (-1,-1) rectangle (1,1);
\filldraw (-1,0) circle (0.1);
 \node [anchor = east] at (-1,0) {$\sigma^c_L$};
\filldraw (1,0) circle (0.1);
 \node [anchor = west] at (1,0) {$\sigma^c_R$};
 \filldraw (0,1) ellipse (0.2 and 0.1);
  \node [anchor = south] at (0,1) {$\bar{M}_c$};
  \filldraw (0,-1) ellipse (0.2 and 0.1);
    \node [anchor = north] at (0,-1) {$M_c$};
\draw [thick]  (0.05,-1)--(0.05,1);
\draw [thick] (-0.05,-1)--(-0.05,1);
\draw[thick, fill=orange] (-0.2,-0.2) rectangle (0.2,0.2);
 \node  at (0.5,0) {$C_\alpha$};
     \filldraw[gray] (-0.75,-1) circle (0.075);
    \node [anchor = south] at (-0.75,-1) {$y_{\bm{g}}$};
\end{tikzpicture}
=
\begin{tikzpicture}[scale=0.75]
\draw[thick, rounded corners, yellow!50!orange] (-1,-1) rectangle (1,1);
\filldraw (-1,0) circle (0.1);
 \node [anchor = east] at (-1,0) {$\sigma^{c'}_L$};
\filldraw (1,0) circle (0.1);
 \node [anchor = west] at (1,0) {$\sigma^{c'}_R$};
 \filldraw (0,1) ellipse (0.2 and 0.1);
  \node [anchor = south] at (0,1) {$\bar{M}_{c'}$};
  \filldraw (0,-1) ellipse (0.2 and 0.1);
    \node [anchor = north] at (0,-1) {$M_{c'}$};
\draw [thick]  (0.05,-1)--(0.05,1);
\draw [thick] (-0.05,-1)--(-0.05,1);
\draw[thick, fill=orange] (-0.2,-0.2) rectangle (0.2,0.2);
 \node  at (0.5,0) {$C_\alpha$};
     \filldraw[gray] (-0.75,-1) circle (0.075);
    \node [anchor = south] at (-0.75,-1) {$y_{\bm{g}}$};
\end{tikzpicture},
\tag{I'}\label{hyp1p:nonSB}
\end{equation}
where $C_\alpha$ (orange square) corresponds to the charge part of the anyon and the operator $y_{\bm{g}}$ (grey circle), defined in \cref{fig:proofobjts}(d), corresponds to the flux part. Following Ref. \cite{Duivenvoorden17}, the condition for an anyon $[g,\alpha]$ to be condensed is
\begin{equation}
\begin{tikzpicture}[scale=0.75]
\draw[thick, rounded corners, yellow!50!orange] (-1,-1) rectangle (1,1);
\filldraw (-1,0) circle (0.1);
 \node [anchor = east] at (-1,0) {$\sigma^c_L$};
\filldraw (1,0) circle (0.1);
 \node [anchor = west] at (1,0) {$\sigma^c_R$};
 \filldraw (0,1) ellipse (0.2 and 0.1);
  \node [anchor = south] at (0,1) {$\bar{M}_c$};
  \filldraw (0,-1) ellipse (0.2 and 0.1);
    \node [anchor = north] at (0,-1) {$M_c$};
\draw [thick]  (0.05,-1)--(0.05,1);
\draw [thick] (-0.05,-1)--(-0.05,1);
\draw[thick, fill=orange] (-0.2,-0.2) rectangle (0.2,0.2);
 \node  at (0.5,0) {$C_\alpha$};
     \filldraw[gray] (-0.75,-1) circle (0.075);
    \node [anchor = south] at (-0.75,-1) {$y_{\bm{g}}$};
\end{tikzpicture} 
\neq 0 \tag{II}\label{hy2:charcond}
\end{equation} 

\

\subsubsection{ SF case}

In this case, we focus on global symmetries that do not permute the anyons, that is
$$
[v_q,u_g]=0 \; \Rightarrow [y_g,x_q]=0 \;  \; \forall q\in Q, g\in G
$$
and that there is a non-trivial SF pattern
\begin{equation}
 v_k v_q = u_{\omega(k,q)} \; v_{kq} \; {\rm with \; some \; } \omega(k,q)\neq 1. \tag{III}\label{hy3:SF}
\end{equation}
We notice that whenever \eqref{hy3:SF} holds, there must exist a charge $\alpha$ that fractionalizes the symmetry:
\begin{equation} 
u_{\omega(k,q)} C_\alpha u^{-1}_{\omega(k,q)} = \chi_\alpha(\omega(k,q)) C_\alpha, \; {\rm s.t.}\; \chi_\alpha(\omega(k,q))\neq 1 \tag{III'} \label{hy3p:SF}
\end{equation}
this is because there must exist a charge $\alpha$ that braids non-trivially the flux labelled by $\omega(k,q)\in G$.

Let us now suppose that \eqref{hy3:SF}, \eqref{hy1:nonSB},  and \eqref{hy2:charcond} hold and that the anyon that fractionalizes the symmetry in \eqref{hy3p:SF}, concretely its charge part, is the one that condenses, satisfying \eqref{hy2:charcond}. Then,
\begin{widetext}
\begin{align*}
\begin{tikzpicture}[scale=0.75]
\draw[thick, rounded corners, yellow!50!orange] (-1,-1) rectangle (1,1);
\filldraw (-1,0) circle (0.1);
 \node [anchor = west] at (-1,0) {$\sigma^c_L$};
\filldraw (1,0) circle (0.1);
 \node [anchor = west] at (1,0) {$\sigma^c_R$};
 \filldraw (0,1) ellipse (0.2 and 0.1);
  \node [anchor = south] at (0,1) {$\bar{M}_c$};
  \filldraw (0,-1) ellipse (0.2 and 0.1);
    \node [anchor = north] at (0,-1) {$M_c$};
\draw [thick]  (0.05,-1)--(0.05,1);
\draw [thick] (-0.05,-1)--(-0.05,1);
\draw[thick, fill=orange] (-0.2,-0.2) rectangle (0.2,0.2);
 \node  at (0.5,0) {$C_\alpha$};
     \filldraw[gray] (-0.75,-1) circle (0.075);
    \node [anchor = south] at (-0.75,-1) {$y_{\bm{g}}$};
\end{tikzpicture} 
& =
\begin{tikzpicture}[scale=0.75]
\draw[thick, rounded corners, yellow!50!orange] (-1,-1) rectangle (1,1);
\filldraw (-1,0) circle (0.1);
 \node [anchor = west] at (-1,0) {$\sigma^c_L$};
\filldraw (1,0) circle (0.1);
 \node [anchor = west] at (1,0) {$\sigma^c_R$};
 \filldraw (0,1) ellipse (0.2 and 0.1);
  \node [anchor = south] at (0,1) {$\bar{M}_{c'}$};
  \filldraw (0,-1) ellipse (0.2 and 0.1);
    \node [anchor = north] at (0,-1) {$M_{c'}$};
\draw [thick]  (0.05,-1)--(0.05,1);
\draw [thick] (-0.05,-1)--(-0.05,1);
    \filldraw[gray] (0.6,1) circle (0.075);
  \node [anchor = south] at (0.6,1) {$x_{q}$};
  \filldraw[gray] (-0.6,1) circle (0.075);
    \node [anchor = south] at (-0.6,1) {$x^{-1}_{q}$};
 \filldraw[gray] (0.6,-1) circle (0.075);
  \node [anchor = north] at (0.6,-1) {$x^{-1}_{q}$};
  \filldraw[gray] (-0.45,-1) circle (0.075);
    \node [anchor = north] at (-0.5,-1) {$x_{q}$};
    \filldraw[gray] (-0.75,-1) circle (0.075);
    \node [anchor = south] at (-0.75,-1) {$y_{\bm{g}}$};
   \filldraw[blue] (0,0.5) ellipse (0.15 and 0.075);
   \node  at (-0.5,0.5) {$\bm{v}_{q}$};
     \filldraw[blue] (0,-0.5) ellipse (0.15 and 0.075);
   \node  at (0.6,-0.5) {$\bm{v}^{-1}_{q}$}; 
\draw[thick, fill=orange] (-0.2,-0.2) rectangle (0.2,0.2);
 \node  at (0.5,0) {$C_\alpha$};
\end{tikzpicture} 
=
\begin{tikzpicture}[scale=0.75]
\draw[thick, rounded corners, yellow!50!orange] (-1,-1) rectangle (1,1);
\filldraw (-1,0) circle (0.1);
 \node [anchor = west] at (-1,0) {$\sigma^c_L$};
\filldraw (1,0) circle (0.1);
 \node [anchor = west] at (1,0) {$\sigma^c_R$};
 \filldraw (0,1) ellipse (0.2 and 0.1);
  \node [anchor = south] at (0,1) {$\bar{M}_{c'}$};
  \filldraw (0,-1) ellipse (0.2 and 0.1);
    \node [anchor = north] at (0,-1) {$M_{c'}$};
\draw [thick]  (0.05,-1)--(0.05,1);
\draw [thick] (-0.05,-1)--(-0.05,1);
    \filldraw[gray] (0.6,1) circle (0.075);
  \node [anchor = south] at (0.6,1) {$x_{q}$};
  \filldraw[gray] (-0.6,1) circle (0.075);
    \node [anchor = south] at (-0.6,1) {$x^{-1}_{q}$};
 \filldraw[gray] (0.6,-1) circle (0.075);
  \node [anchor = north] at (0.6,-1) {$x^{-1}_{q}$};
  \filldraw[gray] (-0.45,-1) circle (0.075);
    \node [anchor = north] at (-0.5,-1) {$ y_{\bm{g}} $};
    \filldraw[gray] (-0.75,-1) circle (0.075);
    \node [anchor = south] at (-0.7,-1) {$x_{q}$};
   \filldraw[blue] (0,0.5) ellipse (0.15 and 0.075);
   \node  at (-0.5,0.5) {$\bm{v}_{q}$};
     \filldraw[blue] (0,-0.5) ellipse (0.15 and 0.075);
   \node  at (0.6,-0.5) {$\bm{v}^{-1}_{q}$}; 
\draw[thick, fill=orange] (-0.2,-0.2) rectangle (0.2,0.2);
 \node  at (0.5,0) {$C_\alpha$};
\end{tikzpicture} 
=
\begin{tikzpicture}[scale=0.75]
\draw[thick, rounded corners, yellow!50!orange] (-1,-1) rectangle (1,1);
\filldraw (-1,0) circle (0.1);
 \node [anchor = west] at (-1,0) {$\sigma^{c'}_L$};
\filldraw (1,0) circle (0.1);
 \node [anchor = west] at (1,0) {$\sigma^{c'}_R$};
 \filldraw (0,1) ellipse (0.2 and 0.1);
  \node [anchor = south] at (0,1) {$\bar{M}_{c'}$};
  \filldraw (0,-1) ellipse (0.2 and 0.1);
    \node [anchor = north] at (0,-1) {$M_{c'}$};
\draw [thick]  (0.05,-1)--(0.05,1);
\draw [thick] (-0.05,-1)--(-0.05,1);
    \filldraw[gray] (-0.75,-1) circle (0.075);
    \node [anchor = south] at (-0.75,-1) {$y_{\bm{g}}$};
   \filldraw[blue] (0,0.5) ellipse (0.15 and 0.075);
   \node  at (-0.5,0.5) {$\bm{v}_{q}$};
     \filldraw[blue] (0,-0.5) ellipse (0.15 and 0.075);
   \node  at (0.6,-0.5) {$\bm{v}^{-1}_{q}$}; 
\draw[thick, fill=orange] (-0.2,-0.2) rectangle (0.2,0.2);
 \node  at (0.5,0) {$C_\alpha$};
\end{tikzpicture} 
=
\begin{tikzpicture}[scale=0.75]
\draw[thick, rounded corners, yellow!50!orange] (-1,-1) rectangle (1,1);
\filldraw (-1,0) circle (0.1);
 \node [anchor = west] at (-1,0) {$\sigma^{c'}_L$};
\filldraw (1,0) circle (0.1);
 \node [anchor = west] at (1,0) {$\sigma^{c'}_R$};
 \filldraw (0,1) ellipse (0.2 and 0.1);
  \node [anchor = south] at (0,1) {$\bar{M}_{c''}$};
  \filldraw (0,-1) ellipse (0.2 and 0.1);
    \node [anchor = north] at (0,-1) {$M_{c''}$};
\draw [thick]  (0.05,-1)--(0.05,1);
\draw [thick] (-0.05,-1)--(-0.05,1);
  \filldraw[blue] (0,0.75) ellipse (0.15 and 0.075);
   \node  at (-0.4,0.75) {$\bm{v}_k$};
  \filldraw[blue] (0,0.5) ellipse (0.15 and 0.075);
   \node  at (-0.4,0.5) {$\bm{v}_q$};
    \filldraw[gray] (0.6,1) circle (0.075);
  \node [anchor = south] at (0.6,1) {$x_k$};
  \filldraw[gray] (-0.6,1) circle (0.075);
    \node [anchor = south] at (-0.6,1) {$x^{-1}_k$};
  \filldraw[blue] (0,-0.75) ellipse (0.15 and 0.075);
   \node  at (0.5,-0.75) {$\bm{v}^{-1}_k$};
     \filldraw[blue] (0,-0.5) ellipse (0.15 and 0.075);
   \node  at (-0.4,-0.5) {$\bm{v}^{-1}_q$}; 
 \filldraw[gray] (0.6,-1) circle (0.075);
  \node [anchor = north] at (0.6,-1) {$x^{-1}_k$};
 \filldraw[gray] (-0.45,-1) circle (0.075);
    \node [anchor = north] at (-0.5,-1) {$x_{k}$};
    \filldraw[gray] (-0.75,-1) circle (0.075);
    \node [anchor = south] at (-0.75,-1) {$y_{\bm{g}}$};
\draw[thick, fill=orange] (-0.2,-0.2) rectangle (0.2,0.2);
 \node  at (0.5,0) {$C_\alpha$};
\end{tikzpicture} 
=
\begin{tikzpicture}[scale=0.75]
\draw[thick, rounded corners, yellow!50!orange] (-1,-1) rectangle (1,1);
\filldraw (-1,0) circle (0.1);
 \node [anchor = west] at (-1,0) {$\sigma^{c''}_L$};
\filldraw (1,0) circle (0.1);
 \node [anchor = west] at (1,0) {$\sigma^{c''}_R$};
 \filldraw (0,1) ellipse (0.2 and 0.1);
  \node [anchor = south] at (0,1) {$\bar{M}_{c''}$};
  \filldraw (0,-1) ellipse (0.2 and 0.1);
    \node [anchor = north] at (0,-1) {$M_{c''}$};
\draw [thick]  (0.05,-1)--(0.05,1);
\draw [thick] (-0.05,-1)--(-0.05,1);
  \filldraw[blue] (0,0.75) ellipse (0.15 and 0.075);
   \node  at (-0.4,0.75) {$\bm{v}_k$};
  \filldraw[blue] (0,0.5) ellipse (0.15 and 0.075);
   \node  at (-0.4,0.5) {$\bm{v}_q$};
  \filldraw[blue] (0,-0.75) ellipse (0.15 and 0.075);
   \node  at (0.5,-0.75) {$\bm{v}^{-1}_k$};
     \filldraw[blue] (0,-0.5) ellipse (0.15 and 0.075);
   \node  at (-0.4,-0.5) {$\bm{v}^{-1}_q$}; 
    \filldraw[gray] (-0.75,-1) circle (0.075);
    \node [anchor = south] at (-0.75,-1) {$y_{\bm{g}}$};
\draw[thick, fill=orange] (-0.2,-0.2) rectangle (0.2,0.2);
 \node  at (0.5,0) {$C_\alpha$};
\end{tikzpicture} 
\\
&
=
\begin{tikzpicture}[scale=0.75]
\draw[thick, rounded corners, yellow!50!orange] (-1,-1) rectangle (1,1);
\filldraw (-1,0) circle (0.1);
 \node [anchor = west] at (-1,0) {$\sigma^{c''}_L$};
\filldraw (1,0) circle (0.1);
 \node [anchor = west] at (1,0) {$\sigma^{c''}_R$};
 \filldraw (0,1) ellipse (0.2 and 0.1);
  \node [anchor = south] at (0,1) {$\bar{M}_{c''}$};
  \filldraw (0,-1) ellipse (0.2 and 0.1);
    \node [anchor = north] at (0,-1) {$M_{c''}$};
\draw [thick]  (0.05,-1)--(0.05,1);
\draw [thick] (-0.05,-1)--(-0.05,1);
  \filldraw[blue] (0,0.75) ellipse (0.15 and 0.075);
   \node  at (-0.4,0.75) {$\bm{v}_{kq}$};
  \filldraw[blue] (0,0.5) ellipse (0.15 and 0.075);
   \node  at (-0.5,0.5) {$\bm{u}_{\omega(k,q)}$};
  \filldraw[blue] (0,-0.75) ellipse (0.15 and 0.075);
   \node  at (0.5,-0.75) {$\bm{v}^{-1}_{kq}$};
     \filldraw[blue] (0,-0.5) ellipse (0.15 and 0.075);
   \node  at (-0.5,-0.5) {$\bm{u}^{-1}_{\omega(k,q)}$}; 
     \filldraw[gray] (-0.75,-1) circle (0.075);
    \node [anchor = south] at (-0.7,-1) {$y_{\bm{g}}$};
\draw[thick, fill=orange] (-0.2,-0.2) rectangle (0.2,0.2);
 \node  at (0.5,0) {$C_\alpha$};
\end{tikzpicture} 
=
\begin{tikzpicture}[scale=0.75]
\draw[thick, rounded corners, yellow!50!orange] (-1,-1) rectangle (1,1);
\filldraw (-1,0) circle (0.1);
 \node [anchor = west] at (-1,0) {$\sigma^{c''}_L$};
\filldraw (1,0) circle (0.1);
 \node [anchor = west] at (1,0) {$\sigma^{c''}_R$};
 \filldraw (0,1) ellipse (0.2 and 0.1);
  \node [anchor = south] at (0,1) {$\bar{M}_{c'''}$};
  \filldraw (0,-1) ellipse (0.2 and 0.1);
    \node [anchor = north] at (0,-1) {$M_{c'''}$};
\draw [thick]  (0.05,-1)--(0.05,1);
\draw [thick] (-0.05,-1)--(-0.05,1);
    \filldraw[gray] (0.6,1) circle (0.075);
  \node [anchor = south] at (0.6,1) {$x^{-1}_{qk}$};
  \filldraw[gray] (-0.6,1) circle (0.075);
    \node [anchor = south] at (-0.6,1) {$x_{qk}$};
 \filldraw[gray] (0.6,-1) circle (0.075);
  \node [anchor = north] at (0.6,-1) {$x_{qk}$};
  \filldraw[gray] (-0.45,-1) circle (0.075);
    \node [anchor = north] at (-0.5,-1) { $x^{-1}_{qk}$ };
    \filldraw[gray] (-0.75,-1) circle (0.075);
    \node [anchor = south] at (-0.7,-1) {$y_{\bm{g}}$};
   \filldraw[blue] (0,0.5) ellipse (0.15 and 0.075);
   \node  at (-0.5,0.5) {$\bm{u}_{\omega(k,q)}$};
     \filldraw[blue] (0,-0.5) ellipse (0.15 and 0.075);
   \node  at (0.6,-0.5) {$\bm{u}^{-1}_{\omega(k,q)}$}; 
\draw[thick, fill=orange] (-0.2,-0.2) rectangle (0.2,0.2);
 \node  at (0.5,0) {$C_\alpha$};
\end{tikzpicture} 
=
\begin{tikzpicture}[scale=0.75]
\draw[thick, rounded corners, yellow!50!orange] (-1,-1) rectangle (1,1);
\filldraw (-1,0) circle (0.1);
 \node [anchor = west] at (-1,0) {$\sigma^{c'''}_L$};
\filldraw (1,0) circle (0.1);
 \node [anchor = west] at (1,0) {$\sigma^{c'''}_R$};
 \filldraw (0,1) ellipse (0.2 and 0.1);
  \node [anchor = south] at (0,1) {$\bar{M}_{c'''}$};
  \filldraw (0,-1) ellipse (0.2 and 0.1);
    \node [anchor = north] at (0,-1) {$M_{c'''}$};
\draw [thick]  (0.05,-1)--(0.05,1);
\draw [thick] (-0.05,-1)--(-0.05,1);
    \filldraw[gray] (-0.75,-1) circle (0.075);
    \node [anchor = south] at (-0.75,-1) {$y_{\bm{g}}$};
   \filldraw[blue] (0,0.5) ellipse (0.15 and 0.075);
   \node  at (-0.5,0.5) {$\bm{u}_{\omega(k,q)}$};
     \filldraw[blue] (0,-0.5) ellipse (0.15 and 0.075);
   \node  at (0.6,-0.5) {$\bm{u}^{-1}_{\omega(k,q)}$}; 
\draw[thick, fill=orange] (-0.2,-0.2) rectangle (0.2,0.2);
 \node  at (0.5,0) {$C_\alpha$};
\end{tikzpicture}
=
\chi_\alpha(\omega(k,q))
\begin{tikzpicture}[scale=0.75]
\draw[thick, rounded corners, yellow!50!orange] (-1,-1) rectangle (1,1);
\filldraw (-1,0) circle (0.1);
 \node [anchor = east] at (-1,0) {$\sigma^{c'''}_L$};
\filldraw (1,0) circle (0.1);
 \node [anchor = west] at (1,0) {$\sigma^{c'''}_R$};
 \filldraw (0,1) ellipse (0.2 and 0.1);
  \node [anchor = south] at (0,1) {$\bar{M}_{c'''}$};
  \filldraw (0,-1) ellipse (0.2 and 0.1);
    \node [anchor = north] at (0,-1) {$M_{c'''}$};
\draw [thick]  (0.05,-1)--(0.05,1);
\draw [thick] (-0.05,-1)--(-0.05,1);
\draw[thick, fill=orange] (-0.2,-0.2) rectangle (0.2,0.2);
 \node  at (0.5,0) {$C_\alpha$};
      \filldraw[gray] (-0.75,-1) circle (0.075);
    \node [anchor = south] at (-0.75,-1) {$y_{\bm{g}}$};
\end{tikzpicture}.
\end{align*}
\end{widetext}
We use now \eqref{hyp1p:nonSB}, arriving to the following equation:
$$
\begin{tikzpicture}[scale=0.75]
\draw[thick, rounded corners, yellow!50!orange] (-1,-1) rectangle (1,1);
\filldraw (-1,0) circle (0.1);
 \node [anchor = east] at (-1,0) {$\sigma^{c}_L$};
\filldraw (1,0) circle (0.1);
 \node [anchor = west] at (1,0) {$\sigma^{c}_R$};
 \filldraw (0,1) ellipse (0.2 and 0.1);
  \node [anchor = south] at (0,1) {$\bar{M}_{c}$};
  \filldraw (0,-1) ellipse (0.2 and 0.1);
    \node [anchor = north] at (0,-1) {$M_{c}$};
\draw [thick]  (0.05,-1)--(0.05,1);
\draw [thick] (-0.05,-1)--(-0.05,1);
\draw[thick, fill=orange] (-0.2,-0.2) rectangle (0.2,0.2);
 \node  at (0.5,0) {$C_\alpha$};
      \filldraw[gray] (-0.75,-1) circle (0.075);
    \node [anchor = south] at (-0.75,-1) {$y_{\bm{g}}$};
\end{tikzpicture}
=
\chi_\alpha(\omega(k,q))
\begin{tikzpicture}[scale=0.75]
\draw[thick, rounded corners, yellow!50!orange] (-1,-1) rectangle (1,1);
\filldraw (-1,0) circle (0.1);
 \node [anchor = east] at (-1,0) {$\sigma^{c}_L$};
\filldraw (1,0) circle (0.1);
 \node [anchor = west] at (1,0) {$\sigma^{c}_R$};
 \filldraw (0,1) ellipse (0.2 and 0.1);
  \node [anchor = south] at (0,1) {$\bar{M}_{c}$};
  \filldraw (0,-1) ellipse (0.2 and 0.1);
    \node [anchor = north] at (0,-1) {$M_{c}$};
\draw [thick]  (0.05,-1)--(0.05,1);
\draw [thick] (-0.05,-1)--(-0.05,1);
\draw[thick, fill=orange] (-0.2,-0.2) rectangle (0.2,0.2);
 \node  at (0.5,0) {$C_\alpha$};
      \filldraw[gray] (-0.75,-1) circle (0.075);
    \node [anchor = south] at (-0.75,-1) {$y_{\bm{g}}$};
\end{tikzpicture}
$$
where we find a contradiction between \eqref{hy3p:SF} and  \eqref{hy2:charcond}.
This means that if a perturbation that preserves the global symmetry also induces an anyon condensation of $[g,\alpha]$ whose charge part $\alpha$ also fractionalizes the symmetry, then the symmetry has to be spontaneously broken after the phase transition in the final phase.

\subsubsection{Permutation case}
In the case where there is a permutation of anyons the symmetry acts on $u_g$ as follows:
$$ \Phi_q(u_g) = v_q u_g v^{-1}_q = u_{\varphi_q(g)}\; {\rm where} \; \varphi_q\in {\rm Aut}(G),$$
where this corresponds to a permutation of the fluxes. Similarly, the permutation of charges is: $\Phi_q(C_\sigma) = C_{ \varphi_q(\sigma)}$ where $\varphi_q(\sigma)$ is another irrep of $G$. We remark that for $G$ abelian, the irreps form a group isomorphic to $G$ so $\varphi_q\in {\rm Aut}(G)$ also characterizes the permutation of the irreps. Generally, for dyons, the transformation is $[g,\alpha]\mapsto [\varphi_q(g),\varphi_q(\alpha)]$. 

If we assume \eqref{hy1:nonSB} that there is no symmetry breaking, \eqref{hy2:charcond} that $[g,\alpha]$ is condensing and that the symmetry permute $[g,\alpha]$ as explained before we obtain that
$$ 
0 \neq
\begin{tikzpicture}[scale=0.75]
\draw[thick, rounded corners, yellow!50!orange] (-1,-1) rectangle (1,1);
\filldraw (-1,0) circle (0.1);
 \node [anchor = west] at (-1,0) {$\sigma^c_L$};
\filldraw (1,0) circle (0.1);
 \node [anchor = west] at (1,0) {$\sigma^c_R$};
 \filldraw (0,1) ellipse (0.2 and 0.1);
  \node [anchor = south] at (0,1) {$\bar{M}_c$};
  \filldraw (0,-1) ellipse (0.2 and 0.1);
    \node [anchor = north] at (0,-1) {$M_c$};
\draw [thick]  (0.05,-1)--(0.05,1);
\draw [thick] (-0.05,-1)--(-0.05,1);
\draw[thick, fill=orange] (-0.2,-0.2) rectangle (0.2,0.2);
 \node  at (0.5,0) {$C_\alpha$};
     \filldraw[gray] (-0.75,-1) circle (0.075);
    \node [anchor = south] at (-0.75,-1) {$y_{\bm{g}}$};
\end{tikzpicture} 
=
\begin{tikzpicture}[scale=0.75]
\draw[thick, rounded corners, yellow!50!orange] (-1,-1) rectangle (1,1);
\filldraw (-1,0) circle (0.1);
 \node [anchor = west] at (-1,0) {$\sigma^c_L$};
\filldraw (1,0) circle (0.1);
 \node [anchor = west] at (1,0) {$\sigma^c_R$};
 \filldraw (0,1) ellipse (0.2 and 0.1);
  \node [anchor = south] at (0,1) {$\bar{M}_{c'}$};
  \filldraw (0,-1) ellipse (0.2 and 0.1);
    \node [anchor = north] at (0,-1) {$M_{c'}$};
\draw [thick]  (0.05,-1)--(0.05,1);
\draw [thick] (-0.05,-1)--(-0.05,1);
    \filldraw[gray] (0.6,1) circle (0.075);
  \node [anchor = south] at (0.6,1) {$x^{-1}_{q}$};
  \filldraw[gray] (-0.6,1) circle (0.075);
    \node [anchor = south] at (-0.6,1) {$x_{q}$};
 \filldraw[gray] (0.6,-1) circle (0.075);
  \node [anchor = north] at (0.6,-1) {$x_{q}$};
  \filldraw[gray] (-0.45,-1) circle (0.075);
    \node [anchor = north] at (-0.5,-1) {$x^{-1}_{q}$};
    \filldraw[gray] (-0.75,-1) circle (0.075);
    \node [anchor = south] at (-0.75,-1) {$y_{\bm{g}}$};
   \filldraw[blue] (0,0.5) ellipse (0.15 and 0.075);
   \node  at (-0.5,0.5) {$\bm{v}_{q}$};
     \filldraw[blue] (0,-0.5) ellipse (0.15 and 0.075);
   \node  at (0.6,-0.5) {$\bm{v}^{-1}_{q}$}; 
\draw[thick, fill=orange] (-0.2,-0.2) rectangle (0.2,0.2);
 \node  at (0.5,0) {$C_\alpha$};
\end{tikzpicture} 
=
\begin{tikzpicture}[scale=0.75]
\draw[thick, rounded corners, yellow!50!orange] (-1,-1) rectangle (1,1);
\filldraw (-1,0) circle (0.1);
 \node [anchor = west] at (-1,0) {$\sigma^{c'}_L$};
\filldraw (1,0) circle (0.1);
 \node [anchor = west] at (1,0) {$\sigma^{c'}_R$};
 \filldraw (0,1) ellipse (0.2 and 0.1);
  \node [anchor = south] at (0,1) {$\bar{M}_{c'}$};
  \filldraw (0,-1) ellipse (0.2 and 0.1);
    \node [anchor = north] at (0,-1) {$M_{c'}$};
\draw [thick]  (0.05,-1)--(0.05,1);
\draw [thick] (-0.05,-1)--(-0.05,1);
\draw[thick, fill=orange] (-0.2,-0.2) rectangle (0.2,0.2);
 \node  at (0.55,0.3) {$C_{\varphi_q(\alpha)}$};
     \filldraw[gray] (-0.5,-1) circle (0.075);
    \node [anchor = south] at (-0.5,-1) {$y_{\varphi_q(\bm{g})}$};
\end{tikzpicture} 
.$$
This implies that the set of condensed anyons are closed under the symmetry. Therefore, if a perturbation that preserves the global symmetry induces an anyon condensation of $[g,\alpha]$ and this is permuted to a non-condensed anyon $[\varphi_q(g),\varphi_q(\alpha)]$, the symmetry has to be spontaneously broken after the phase transition in the final phase.

\section{A toric code state with a $\mathbb{Z}_2$ global symmetry on the square lattice}  \label{app:TCPEPS}

In this appendix, we consider deformations of a fixed-point tensor representing a toric code enriched with $\mathbb{Z}_2$ symmetry in order to test that the string order parameters of Eq.~\ref{SOPtnl} function as intended away from the fixed-point.

We use the following PEPS tensor to model the ground state of the symmetry-enriched toric code,%$$A= \sum_{b=0,1} {(X^2/\sqrt{2})^b}^{\otimes 4},$$
\begin{equation}\label{eq:AX2}
A= \frac{1}{\sqrt{2}}\sum_{b=\{0,1\}} {(X^2)^b}^{\otimes 4}
\end{equation}
%This normalization is requiered for A being a projector
where here, and throughout this section, $X=\sum_{i=0}^3 |i+1 \; (\text{mod}\;4)\rangle \langle i|$
%$X=\left(\begin{array}{cccc}0&1&0&0 \\ 0&0&1&0\\0&0&0&1 \\ 1&0&0&0 \end{array}\right)$ 
is the generator of the left regular representation of $\mathbb{Z}_4$ such that $X^4=\id$ and $X^2=\sigma_x\otimes \id_2\equiv g$. The tensor $A$ is $\mathbb{Z}_2$-invariant, i.e. $A$ has the following virtual symmetry $A=A(g\otimes g\otimes g\otimes g)$. The PEPS constructed with $A$, $|\psi_A\rangle$, is left invariant by the action of the operator $U = X\otimes X\otimes {X}^{-1}\otimes {X}^{-1}$ on each lattice site:  $U^{\otimes N}| \psi_A\rangle=| \psi_A\rangle$ which would correspond to a global $\mathbb{Z}_4$ symmetry. However, the state $|\psi_A\rangle$ has also a local symmetry generated by $U^2$ such that $U^2_i|\psi_A\rangle = |\psi_A\rangle$, where $i$ is any site. Therefore, the {\it a priori} global $\mathbb{Z}_4$ symmetry is reduced to a  global $\mathbb{Z}_2$ symmetry when quotiented by the local symmetry: this is the one we consider here.

The interesting feature of this model, when considering the parent Hamiltonian of $| \psi_A\rangle$, is that the anyonic excitation corresponding to the charge fractionalizes that global $\mathbb{Z}_2$ symmetry. More explicitly, the virtual operator of the charge is  $C_\sigma=\sigma_z \otimes \id_2$ and the virtual action of the global $\mathbb{Z}_2$ symmetry is given by the conjugation with $X$ such that  $C_\sigma\rightarrow X C_\sigma X^{-1}$. Then, if we apply twice the global symmetry an isolated charge changes as $C_\sigma\rightarrow - C_\sigma $ which corresponds to a projective representation of $\mathbb{Z}_2$.

In the following, we study the effects of different perturbations on the wavefunction generated by \eqref{eq:AX2}

\subsection{ Diagonal string perturbation }

We apply the perturbation $({\exp}[\theta X^2/2])^{\otimes 4}$ to the tensor $A$ where ${\exp}[\theta X^2]={\cosh}(\theta)\id+{\sinh}(\theta)X^2$. The perturbation commutes with the symmetry and it drives the system from the TC phase to a product state. This is because $\lim_{\theta \rightarrow \infty}\frac{\sinh(\theta)}{\cosh(\theta)}=1$ so that for large values of $\theta$ we have $ {\exp}[\theta X^2 /2] \propto \id+X^2$. This implies that
$$\lim_{\theta \rightarrow \infty}({\exp}[\theta X^2/2])^{\otimes 4}A \propto (\id+X^2)^{\otimes 4}\equiv A_{\infty}.$$
Since all the operators involved are real the on-site transfer operator can be written as
$$\mathbb{E}=A^*(\theta)A(\theta)= {\exp}[\theta X^2]^{\otimes 4} [\id^{\otimes 4} + (X^2)^{\otimes 4}].$$ 
The norm of the PEPS with the perturbation is calculated with the contraction of this tensor $\mathbb{E}$ on each vertex $v$ of the square lattice $\mathcal{V}$. When two sites, $i$ and $j$,  coincide the resulting factor is 
\begin{align*}L_\theta(b_i,b_j)= & \tr[{\cosh}(2\theta) (X^2)^{b_i-b_j}+{\sinh}(2\theta) (X^2)^{b_i-b_j+1}] \\ = &\bigg\{ \begin{array}{lcc}   
4 \cosh(2 \theta)& {\rm if} & b_i-b_j=0 \; {\rm mod\ 2 }
\\
4 \sinh(2 \theta)&  {\rm if} & b_i-b_j=1 \; {\rm mod\ 2 } \end{array},
\end{align*}
instead of the value $4 \delta_{b_i-b_j,0}$ in the non-perturbed case $\theta=0$.
Then, the norm can be expressed as the following sum
\begin{equation}\label{normstate}
\langle {\psi_{A(\theta)}}|{\psi_{A(\theta)}}\rangle= \mathcal{C}_{v\in \mathcal{V}}\{ \mathbb{E}_v \}=\sum_{\bm{b}}\prod_{\langle i,j \rangle} L_\theta (b_i,b_j),
\end{equation}
so that $\langle {\psi_{A(0)}}|{\psi_{A(0)}}\rangle = 2\cdot 4^{2N_s}$  where $N_s$ is the number of sites of $\mathcal{V}$. The wavefunction $|\psi_{A(\theta)}\rangle$ can be normalized locally by modifying the weight of the tensor $A$.  We find that the string order parameters are $ \mathcal{L}^{[q]}_\ell(0) = (-1)^q\cdot 2 \cdot 4^{2N_s-4}$ at the fixed-point, and 
\begin{align*}
\mathcal{L}^{[a]}_1(\theta)= &  \sum_{\bm{b}}   A(b_6,b_7) B(b_3,b_4,b_5,b_6) C_{b_4,b_5}(b_1,b_2)C_{b_4,b_5}(b_8,b_9),
\end{align*}
and similarly for $\mathcal{L}^{[e]}_1(\theta)$. The values of the functions  $A,B,C$ are
$$  A(b_6,b_7) =\left\{ \begin{array}{lcc}  
0 & {\rm if} & b_7=1,\ b_6=0
\\
-4&  {\rm if} & b_7=1,\ b_6=1
\\
4&  {\rm if} & b_7=0,\ b_6=0
\\
0&  {\rm if} & b_7=0,\ b_6=1
 \end{array} \right. 
 =2(s_6+s_7),
 $$
 
\begin{align*}  B(b_3,b_4, b_5,b_6) &=\left \{
 \begin{array}{lcc}   
4 & {\rm if} & b_3+b_6= 0 , \; b_4+b_5+b_6= 0
\\
0&  {\rm if} & b_3+b_6= 1 , \; b_4+b_5+b_6= 1
\\
0&  {\rm if} & b_3+b_6= 1 , \; b_4+b_5+b_6= 0
\\
-4&  {\rm if} & b_3+b_6= 0 ,  \; b_4+b_5+b_6= 1
\end{array} \right.\ \\
&=2s_4s_5(s_3+s_6),
\end{align*}
and
\begin{align*}   C_{b_4,b_5}(b_i,b_j) &=\left \{
 \begin{array}{lcc}   
2(e^{4\theta}-e^{-4\theta}) & {\rm if} & b_i+b_j+b_4+b_5= 0 
\\
2(e^{4\theta}+e^{-4\theta})  & {\rm if} & b_i+b_j+b_4+b_5= 1
\end{array} \right.\\
&=  2(e^{4\theta}-s_is_js_4s_5e^{-4\theta}),
\end{align*}
where all the sums are modulo 2 and the transformation to spin variables is $s_i=(-1)^{b_i}$. The sublattice $\Omega$ corresponds to the following set of edges that are involved in the SOP:
$$\Omega=\{   \langle 1,4 \rangle, \langle2,5\rangle,\langle3,4\rangle,\langle4,5\rangle,\langle5,6\rangle,  \langle6,7\rangle, \langle4,8\rangle,\langle5,9\rangle\},$$
which are placed as follows:
\begin{equation}\nonumber
\cbox{11.0}{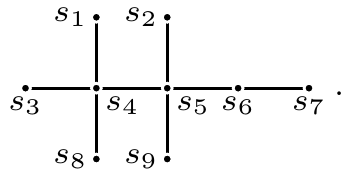}
\end{equation}

It can be checked that for $\theta = 0$, $b_i=b_j$ except for $b_4,b_5$ in $\mathcal{L}^{[a]}_1(0)$, such that we obtain $\mathcal{{O}}_1(0) =  -1$.

\subsubsection*{Mapping to the classical 2D Ising model}

\begin{figure}
\begin{center}
\includegraphics[width=26em]{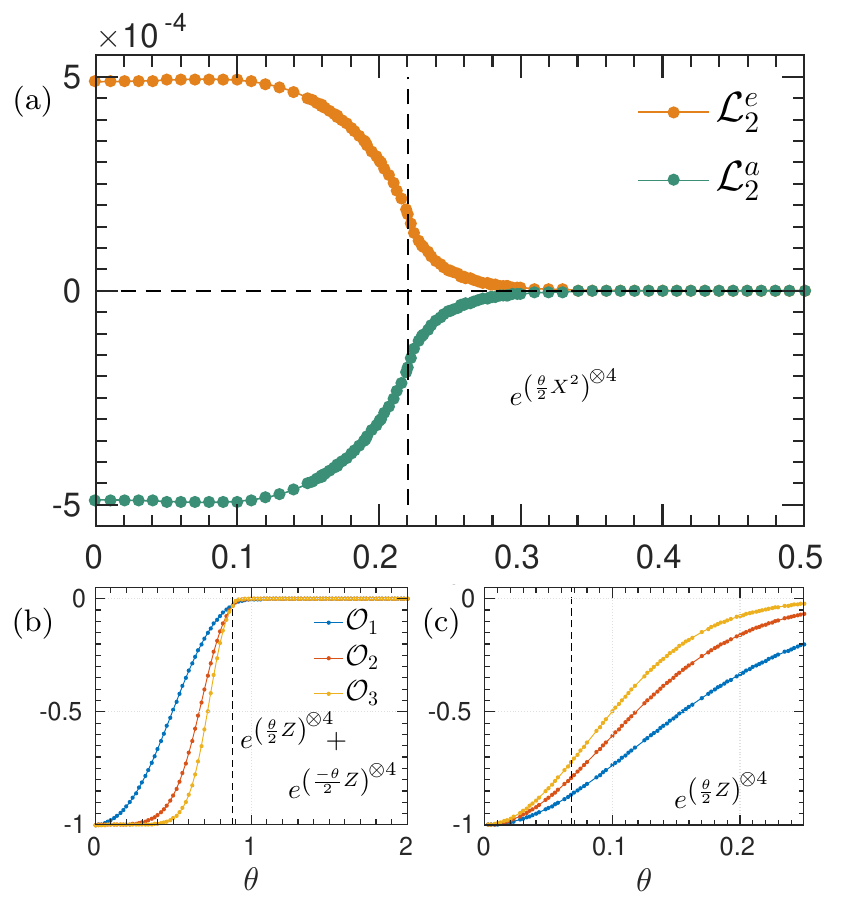}
\caption{ (a) Results for
$\mathcal{L}^{[a]}_2$ and $\mathcal{L}^{[e]}_2$ of the phase transition driven by the perturbation ${\rm exp}[\frac{\theta }{2}X^2]^{\otimes 4} $. $\mathcal{{O}}_\ell(\theta) $ computed for the phase transitions driven  by perturbations 
(b) ${\rm exp}[\frac{\theta}{2} Z]^{\otimes 4} + {\rm exp}[-\frac{\theta}{2} Z]^{\otimes 4}$, and (c) ${\rm exp}[\theta Z]^{\otimes 4} $. All data are obtained by using tensor network states algorithms.}
\label{fig:filterings}
\end{center}
\end{figure}

The norm of Eq. \eqref{normstate} can be written as the partition function of the classical Ising model
\begin{equation} \label{den}
\sum_{\bm{s}}\prod_{ \langle i,j \rangle } e^{\beta s_i s_j},
\end{equation}
where $s_i=(-1)^{b_i}$ are the Ising variables and the different weights correspond to
\begin{equation}\label{transf}
\bigg\{ \begin{array}{lcc}   
e^\beta  &= 4 \cosh(2\theta)&= 2(e^{2\theta}+e^{-2\theta}),
\\
e^{-\beta}  &= 4 \sinh(2\theta)&=2( e^{2\theta}-e^{-2\theta})
\end{array}
\bigg\}
\end{equation}
If we compare the ratio of the above expressions we obtain $e^{-2\beta}=\tanh(2\theta)$. So, the critical temperature $\beta_c= \ln(1+\sqrt{2})/2$ corresponds to $\theta_c=\beta_c/2$ since $\theta=\tanh^{-1}(e^{-2\beta})/2$. 
 
In order to compute Eq. \eqref{SOPtnl}, we need to express its numerator in the spins variables.

The calculation results in the following 
\begin{align} \label{eq:opising}
\mathcal{L}^{[a]}_1(\theta)  \propto &  \sum_{\bm{s}}  (s_6+s_7) s_4s_5(s_3+s_6) (1-s_1s_2s_4s_5e^{-8\theta}) \notag \\ 
& \times
(1-s_8s_9s_4s_5e^{-8\theta})
 e^{8\theta} \prod_{\langle i,j \rangle \in \mathcal{V}-\Omega} e^{\beta s_i s_j} 
\end{align}
where $\beta$ depends on $\theta$ via Eq.\eqref{transf} as $e^{-2\beta}=\tanh(2\theta)$.

We can also map to the Ising model the order parameter for a blocking of $\ell$ sites:
\begin{align*}
\mathcal{L}^{[a]}_\ell(\theta)  &  \propto \sum_{\bm{s}}    \; s_{3\ell+1}s_{3\ell+2}(s_{3\ell+2}+s_{4\ell+2}) (s_{5\ell+1}+s_{5\ell+2})
\\
& \times  \prod_{\substack{ i=1,\ell \\j=0,5\ell +2}} (1-s_{2\ell+1+i}s_{3\ell+1+i}s_{i+j}s_{i+\ell+j}e^{-8\theta}) \\
&\times \prod_{\langle i,j \rangle\in \mathcal{V}-\Omega} e^{\beta s_i s_j} ,
\end{align*}
where $\Omega_\ell$ is now the sublattice composed by the following red edges:
\begin{equation}\nonumber
\cbox{24.5}{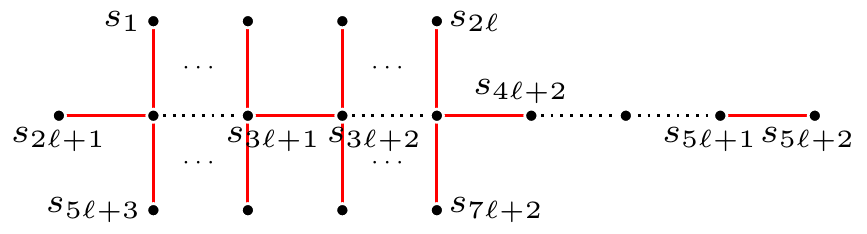}
\end{equation}
In Fig.\ref{fig:filterings}(a) we show the values of $\mathcal{L}^{[a]}_2(\theta)$ and $\mathcal{L}^{[e]}_2(\theta)$. It easy to see that $\mathcal{L}^{[e]}_2(\theta)$ corresponds to the confinement fraction of the charge \cite{Duivenvoorden17,Iqbal18} and it goes to zero after the phase transition. Therefore, \eqref{transprotocol} is no longer valid in that region since the denominator goes to zero. But before the phase transition, we can see that $\mathcal{O}^{[a]}=-1$.

We use the infinite matrix product state (iMPS) algorithm for these computations \cite{haegeman2017diagonalizing}. Furthermore, since, in this case, the deformation keeps the model exactly solvable, we use also use the Metropolis-Hastings algorithm \cite{Hastings70}, which allows us to evaluate $\mathcal{{O}}_\ell(\theta) $ for even larger values of $\ell$ (data not shown).

\subsection{Dual string perturbations}

Now, we analyze the behavior of the string order parameter by considering two further perturbations.

\begin{enumerate}

\item We start by defining $Z=\rm diag(1,-1,1,-1) = \id_2 \otimes \sigma_z$. Since $Z$ commutes with 
$X^2$, by applying the perturbation $P(\theta) = ({\rm exp}[\frac{\theta}{2} Z])^{\otimes 4} + ({\rm exp}[-\frac{\theta}{2} Z])^{\otimes 4}$, the resulting tensor in the limiting case can be given as,
$$\lim_{\theta \rightarrow \infty}P(\theta)A \propto \left( \sigma_+^{\otimes 4} +\sigma_-^{\otimes 4} \right) \otimes \left [\id_2^{\otimes 4} +X^{\otimes 4} \right].$$
The perturbed state corresponds to two copies of regular toric code times product states. Moreover the perturbation commutes with the symmetry, the results of the numerics are shown in Fig. \ref{fig:filterings}(b) where it can be seen that $\mathcal{{O}}_\ell(\theta)$ gets sharper with increasing $\ell$.

\item On the other hand, if we restrict the perturbation to  $P(\theta) = {({\rm exp}[\frac{\theta}{2} Z])}^{\otimes 4}$ the resulting tensor in the limiting case is 
 $$\lim_{\theta \rightarrow \infty}({\rm exp}[\theta Z])^{\otimes 4}A \propto \left[ |0\rangle \langle 0|^{\otimes 4} +|0\rangle \langle 1|^{\otimes 4} \right]   \otimes \id_2^{\otimes 4},$$
which corresponds to a product state. It is important to note that in this example the perturbation doesn't commute with the symmetry so that the SOP is no longer well defined. This is what Fig. \ref{fig:filterings}(c) shows, a slow decay of $\mathcal{{O}}_\ell(\theta)$ in contrast with the other sharp behaviours when the perturbations commute with the symmetry.
\end{enumerate}

\bibliographystyle{apsrev4-1}
\bibliography{main}

\end{document}